\documentclass[fleqn,usenatbib]{mnras}

\usepackage{newtxtext,newtxmath}

\usepackage[T1]{fontenc}

\DeclareRobustCommand{\VAN}[3]{#2}
\let\VANthebibliography\thebibliography
\def\thebibliography{\DeclareRobustCommand{\VAN}[3]{##3}\VANthebibliography}

\usepackage{graphicx}	
\usepackage{amsmath}	
\usepackage{siunitx}
\usepackage{booktabs}
\usepackage{subfloat}

\newcommand{\oii}{[O\,\textsc{ii}]~}
\newcommand{\oiii}{[O\,\textsc{iii}]}
\newcommand{\sii}{[S\,\textsc{ii}]~}
\newcommand{\siii}{[S\,\textsc{iii}]~}
\newcommand{\siv}{[S\,\textsc{iv}]~}
\newcommand{\nii}{[N\,\textsc{ii}]~}

\newcommand{\nv}{N\,\textsc{v}~}
\newcommand{\ha}{H$\rm\alpha$~}
\newcommand{\hb}{H$\rm\beta$~}
\newcommand{\lya}{Ly$\rm\alpha$}

\newcommand{\ciii}{C\,\textsc{iii}]}
\newcommand{\civ}{C\,\textsc{iv}}
\newcommand{\heii}{He\,\textsc{ii}}
\newcommand{\oiiiuv}{O\,\textsc{iii}]}

\newcommand{\cii}{[C\,\textsc{ii}]~}
\newcommand{\neii}{[Ne\,\textsc{ii}]~}
\newcommand{\neiii}{[Ne\,\textsc{iii}]~}

\newcommand{\hii}{H\,\textsc{ii}~}
\newcommand{\fesc}{$\rm f_{esc}$(LyC)}

\newcommand{\lm}{$\lambda$~}
\newcommand{\lmlm}{$\lambda\lambda$~}
\newcommand{\kms}{km s$^{-1}$}
\usepackage{siunitx}

\title[Pox 186]{A study of  extreme CIII]1908 \& [OIII]88/[CII]157 emission in  Pox 186: implications for JWST+ALMA (FUV+FIR) studies of distant galaxies}
\author[Nimisha Kumari et al.]{
Nimisha Kumari$^1$\thanks{E-mail: kumari@stsci.edu},
Renske Smit$^2$,
Claus Leitherer$^3$,
Joris Witstok$^{4,5}$,
Mike J Irwin$^6$,
\and
Marco Sirianni$^7$, 
Alessandra Aloisi$^3$
\\
$^{1}$AURA for European Space Agency (ESA), ESA Office, Space Telescope Science Institute, 3700 San Matin Drive, Baltimore, MD, 21218, USA\\
$^{2}$Astrophysics Research Institute, Liverpool John Moores University, Liverpool, L35 UG, UK\\
$^{3}$Space Telescope Science Institute, Baltimore, MD, 21218, USA\\
$^4$Kavli Institute for Cosmology, University of Cambridge, Madingley Road, Cambridge CB3 0HA, UK\\
$^5$Cavendish Laboratory, University of Cambridge, 19 JJ Thomson Avenue, Cambridge CB3 0HE, UK\\
$^6$Insitute of Astronomy, University of Cambridge, Madingley Road, Cambridge CB3 0HA, UK\\
$^7$European Space Agency (ESA), ESA Office, Space Telescope Insitute, 3700 San Martin Drive, Baltimore, MD 21218, USA}

\date{Accepted XXX. Received YYY; in original form ZZZ}

\pubyear{2023}

\begin{document}
\label{firstpage}
\pagerange{\pageref{firstpage}--\pageref{lastpage}}
\maketitle

\begin{abstract}
Carbon spectral features are ubiquitous in the ultraviolet (UV) and far-infrared (FIR) spectra of galaxies in the epoch of reionization (EoR). We probe the ionized carbon content of a blue compact dwarf galaxy Pox 186 using the UV, optical, mid-infrared and FIR data taken with telescopes in space (Hubble, Spitzer, Herschel) and on the ground (Gemini). This local (z$\sim$0.0040705) galaxy is likely an analogue of EoR galaxies, as revealed by its extreme FIR emission line ratio, \oiii \SI{88}{\micro\meter}/\cii \SI{157}{\micro\meter} ($>$10). The UV spectra reveal  extreme \ciii \lmlm 1907, 1909 emission with the strongest equivalent width (EW) = 35.85 $\pm$ 0.73 \AA~ detected so far in the local (z$\sim$0) Universe, a relatively strong \civ \lmlm 1548, 1550 emission with EW = 7.95 $\pm$0.45\AA, but no \heii\lm 1640 detection.  Several scenarios are explored to explain the high EW of carbon lines, including high effective temperature, high carbon-to-oxygen ratio, slope and upper mass of top-heavy initial mass function, hard ionizing radiation and in-homogeneous dust distribution.   Both \ciii~ and \civ~ line profiles are broadened with respect to the \oiiiuv\lm 1660 emission line. Each emission line of \civ \lmlm 1548, 1550 shows the most distinct double-peak structure ever detected which we model via two scenarios, firstly a double-peaked profile that might emerge from resonant scattering and secondly a single nebular emission line along with a weaker interstellar absorption. The study demonstrates that galaxies with extreme FIR emission line ratio may also show extreme UV properties, hence paving a promising avenue of using FIR+UV in the local (via HST+Herschel/SOFIA) and distant (via JWST+ALMA) Universe for unveiling the mysteries of the EoR.

\end{abstract}

\begin{keywords}
galaxies:dwarfs -- galaxies:abundances -- galaxies:high-redshift
\end{keywords}



\section{Introduction}
\label{section:intro}

\begin{figure}
    \centering
    \includegraphics[width=0.45\textwidth]{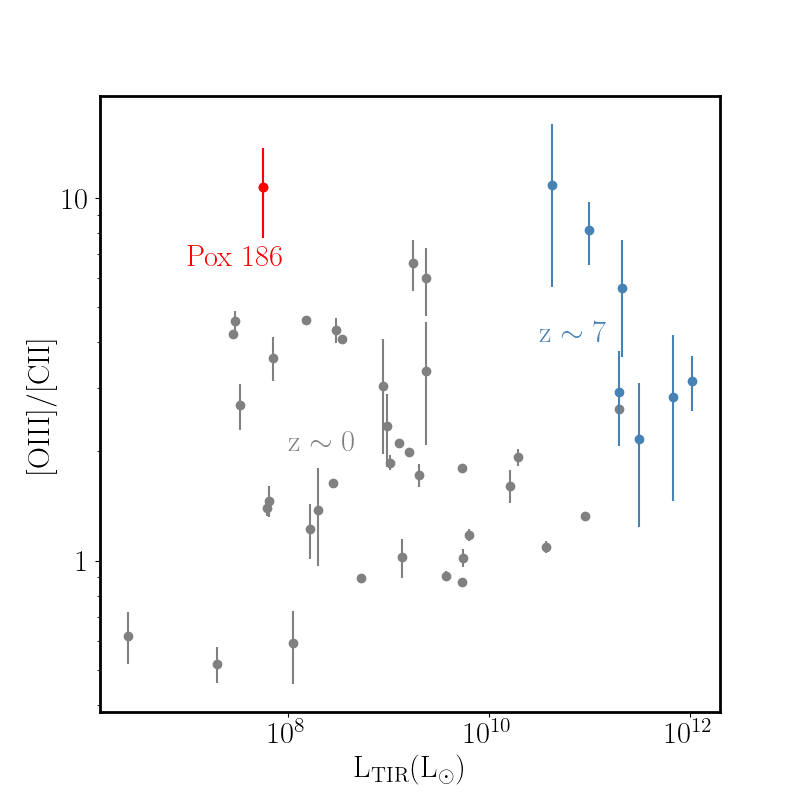}
    \caption{\oiii~ \SI{88}{\micro\meter}/\cii \SI{157}{\micro\meter} line ratio versus IR luminosity for Pox 186 (red point), galaxies at z=7 \citep[blue points,][]{Hashimoto2019,
Witstok2022, Ren2023} and local dwarf galaxies \citep[grey points,][]{Cormier2015}.}
    \label{fig:OIIICII}
\end{figure}

\indent Understanding the reionization of the Universe is one of the frontier goals of modern astronomy. Several theoretical and observational efforts have been made to answer the related pressing questions such as when and how first galaxies formed \citep[e.g.,][]{Stark2016} and whether these first galaxies reionized the intergalactic medium \citep[IGM; e.g.,][]{Robertson2010}. A first step to answer these questions is to search for the early galaxies in the epoch of reionization (EoR) and characterize their properties. 

\indent Deep imaging campaigns have been quite successful in searching for such sources. For example, the Hubble Space Telescope (HST) allowed us to compile large samples of early galaxies via deep near-infrared (NIR) imaging programs such as the Great Observatories Origins Deep Survey \citep[GOODS,][]{Giavalisco2004}, Extreme Deep Field \citep[XDF,][]{Illingworth2013} and Cosmic Assemble Near-Infrared Deep Extragalactic Legacy Survey \citep[CANDELS,][]{Grogin2011, Koekemoer2011}. However, spectroscopy is essential to characterize the properties of these sources.  Until the launch of the James Webb Space Telescope (JWST), the rest-frame ultraviolet (UV) spectroscopy was available for only a few EoR galaxies \citep[e.g.,][]{Sobral2015, Stark2017, Topping2021, Hutchison2019}. 
The JWST observations are significantly improving the dearth of UV spectroscopy for the EoR galaxies via the planned follow-up spectroscopic surveys of the Hubble deep fields \citep{Robertson2021, Curtislake2022, Bunker2023}. Similarly, the exquisite sensitivity of the Atacama Large Millimeter Array (ALMA) has made it possible to obtain the far-infrared (FIR) spectroscopy of EoR galaxies \citep[e.g.,][]{Maiolino2015, Carniani2017, Smit2018, Bouwens2022, Witstok2022}. The combined JWST+ALMA spectroscopic observations will significantly enhance our understanding of the EoR.

\indent An indirect approach to probe the nature of reionization sources while efficiently using the two simultaneously-operating state-of-the-art facilities, JWST and ALMA, is to perform detailed studies of the physical processes operating in local galaxies which might resemble EoR galaxies. Several different criteria have been devised so far to identify the local analogues of high-redshift galaxies, including gas-phase metallicity, star-formation rate, compactness, stellar mass, UV luminosity, dust attenuation, \lya~ emission, colour, and ionization state among many others. Some established classes of local analogues of high-redshift galaxies are blue compact dwarf galaxies \citep[BCD,][]{SearleSargent1972}, green peas \citep[GP,][]{Cardamone2009} and  blueberries \citep{Yang2017}, though it is not clear whether these galaxy populations also resemble the EoR galaxies, mainly because of the dearth of data available on EoR galaxies so far.

\indent One of the goals of this paper is to demonstrate the use of FIR line ratio  \oiii ~\SI{88}{\micro\meter}/\cii \SI{157}{\micro\meter} for identifying the local analogues of the EoR galaxies. \oiii  
 \SI{88}{\micro\meter} originate from the ionized gas, \cii \SI{157}{\micro\meter} may originate from both the ionized as well as neutral interstellar medium (ISM), and their relative strengths (i.e. \oiii ~\SI{88}{\micro\meter}/\cii \SI{157}{\micro\meter} line ratio) can potentially tell us about the porosity of ISM \citep{Chevance2016, Polles2019}. The ALMA observations of EoR galaxies (Figure \ref{fig:OIIICII}, blue points) reveal that  \oiii ~\SI{88}{\micro\meter}/\cii \SI{157}{\micro\meter} line ratio may vary in the range 1--10, indicating a highly porous ISM which will facilitate the leakage of ionizing photons required for reionization of the neutral IGM. The Herschel Dwarf Galaxies Survey \citep{Madden2013, Cormier2015} reveal a large population of local dwarf galaxies with  \oiii ~\SI{88}{\micro\meter}/\cii \SI{157}{\micro\meter} $>$ 1 (Figure \ref{fig:OIIICII}, grey points), which are potentially the local analogues of EoR galaxies.       

\begin{figure}
    \centering
    \includegraphics[width=0.48\textwidth]{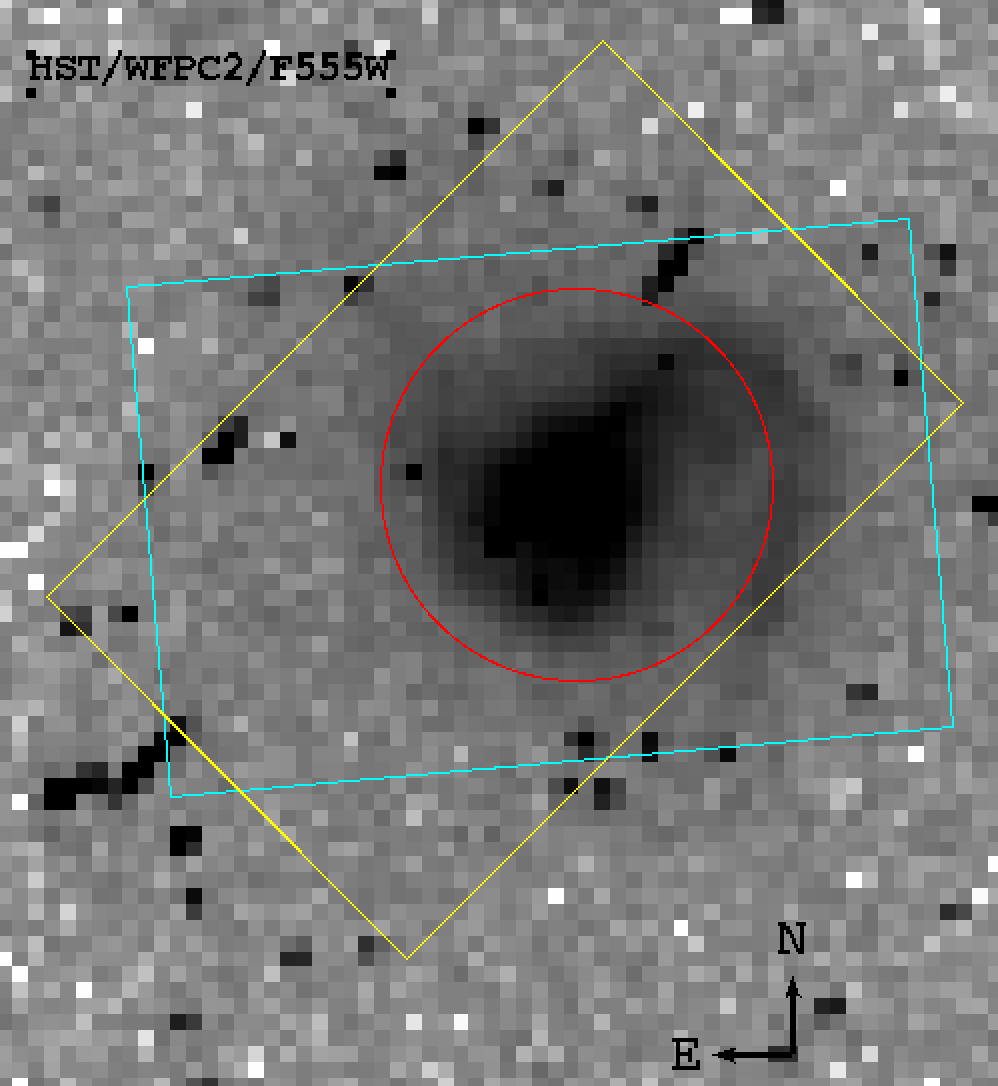}
    \caption{HST/WFPC2 image of Pox 186 taken in F555W filter (Prog id: 8333) on which we overlay FOVs of all primary observations used in this work. The red circle denotes the HST/COS aperture of 1.25$\arcsec$ radius (PIDs: 16445, 16071). The yellow and cyan rectangles of size 3.5$\arcsec\times$5$\arcsec$ denote the FOVs of GMOS-N IFU programs GN-2020A-FT-105 and GN-2021A-FT-111, respectively.}
    \label{fig:fov}
\end{figure}

\indent In an attempt to explore and establish using \oiii ~\SI{88}{\micro\meter}/\cii \SI{157}{\micro\meter} as a criterion for identifying the EoR local analogues, we obtained HST UV and spatially-resolved optical spectroscopy of Pox 186, a unique dwarf galaxy showing the highest \oiii ~\SI{88}{\micro\meter}/\cii \SI{157}{\micro\meter} ever detected in the local Universe (Figure \ref{fig:OIIICII}, red point). Moreover, \citet{Ramambason2022} shows that this galaxy has an ionizing photon escape fraction of $\sim$ 40\%, thus making it ideal for this study. Pox 186 was originally discovered in \citet{Kunth1981}, and was thought to be a protogalaxy. \citet{Corbin2002} later shows that Pox186 is an ultra-compact galaxy still in the process of formation with a majority of star-formation concentrated in the central star cluster of mass 10$^5$ M$_{\odot}$.   
Figure \ref{fig:fov} shows a narrow-band optical image of Pox 186 taken with Wide Field Planetary Camera2 onboard HST, along with the field-of-view (FOV) of instruments of primary observations used in this work. Table \ref{tab:data} lists some of the main physical properties of Pox 186, along with information about the UV and optical observing strategy. A typical UV spectrum of star-forming galaxies is known to show prominent spectral features such as \lya\lm 1215,  \civ \lmlm 1548, 1550, \heii\lm 1640, \oiiiuv \lmlm 1660, 1666,  [C \textsc{iii}]\lm1907 and \ciii\lm1909 \footnote{In the rest of the paper, we will refer the two carbon emission lines [C \textsc{iii}]\lm1907 and \ciii\lm1909 as \ciii\lmlm1907,1909, which is a popular notation in literature.}, which have been used to infer information regarding hardness of radiation fields, ionization conditions, metal-content, wind properties within galaxies at all redshifts \citep[e.g.,][]{Shapley2003, Senchyna2017, Nakajima2018, Schmidt2021}. In this paper, we mainly focus on the ionized carbon spectral features, \civ \lmlm 1548, 1550 and \ciii\lmlm1907,1909, however, we complement the UV analysis with the spatially-resolved optical, mid-infrared (MIR) and FIR data. 

\indent The paper is organized as follows: Section \ref{section:observations} presents an overview of the data used in this work, including UV, optical, MIR and FIR. For UV and optical data, we explain the initial data reduction and processing. The MIR and FIR data are archival. In Section \ref{section:results}, we present the results of the multi-wavelength data analysis which includes the estimates of redshift, distance, flux and equivalent widths of detected emission lines and reddening. We also determine several physical properties of the ionized gas and the ionizing stellar population, such as electron temperature and density, gas-phase metallicity, ionization parameters, effective temperature and softness parameters.  Section \ref{section:discussion} presents a discussion focusing on UV carbon features including their large equivalent widths, line profiles and relative chemical abundance. We also discuss the implication of this study on future JWST+ALMA studies of reionization-era galaxies. Section \ref{sec:summary} summarizes our main results.  

\indent In the rest of the paper, we assume a flat $\Lambda$CDM cosmology with H$_{0}$ = 70 \kms Mpc$^{-1}$, $\Omega_{m}$ = 0.3. The gas-phase solar metallicity is assumed to be 12 + log(O/H)$_{\odot}$ = 8.69 \citep{Asplund2009}.

\begin{table}

\caption{Basic data and HST/COS exposure times for Pox 186}
\resizebox{0.99\columnwidth}{!}
{%
\begin{tabular}{@{}lcr@{}}
\toprule
& \multicolumn{1}{c}{Pox 186}                    \\ \midrule
Morphological type    &     & BCD              \\
RA (J2000)            &     & 13 25 48.641      \\
Dec (J2000)           &     & -11 36 37.94      \\
Redshift$^a$          &   & 0.0040705         \\
Distance (Mpc)$^a$    &     & 17.5              \\
V (mag)$^b$           &       & 17.43 $\pm$ 0.03  \\
B (mag)$^b$           &     & 17.93 $\pm$ 0.53  \\
M$\rm_{HI}$ (M$_{\odot}$)$^c$  & & $<$ 16.1 $\times$ 10$^5$          \\
\midrule
COS Setting                & Central wavelength (\AA) & Exposure time (s) \\
\midrule
G130M$^d$                 & 1291  & 5123.36 \\
G160M$^e$                & 1623 & 4528.064          \\
G185M$^e$                & 1913 & 13428.864         \\
Mirror A                  & -- & 7 $\times$ 3$^f$     \\ 

\midrule
GMOS gratings  &Central wavelength (\AA) &  Exposure time (s) \\
\midrule 
B600+\_G5307$^g$ &4650& 870$\times$3\\
B600+\_G5307$^g$ &4700& 870$\times$3\\
R831+\_G5302$^g$ &6900& 950$\times$2\\
R831+\_G5302$^g$ &6950& 950$\times$3\\
R831+\_G5302$^h$ &8800&1000$\times$3 \\
R831+\_G5302$^h$ &8900&1000$\times$4 \\
\bottomrule
\end{tabular}%
}
\\Notes:
$^a$: This work;
$^b$: \citet{Guseva2004};
$^c$: \citet{Begum2005};
$^d$: HST/COS PID 16445;
$^e$: HST/COS PID 16071;
$^f$: Target acquisition was done thrice using Mirror A, every time before taking COS/FUV and COS/NUV spectra;
$^g$: GMOS PID:GN-2020A-FT-105;
$^h$: GMOS PID: GN-2021A-FT-111.
\label{tab:data}
\end{table}

\begin{table}
\caption{IR emission line fluxes of Pox 186 from \citet{Cormier2015}.}
\resizebox{0.99\columnwidth}{!}{%
    \begin{tabular}{@{}lccccccr@{}}
    \toprule
    Emission lines & & & & & & & Observed Fluxes \\
     \midrule
       \siv \SI{10.5}{\micro\meter} & & & & & & & 36.29 $\pm$ 3.66 \\
        \neii \SI{12.8}{\micro\meter} & & & & & & & 0.95 $\pm$ 0.25 \\
          \neiii \SI{15.6}{\micro\meter} & & & & & & &   13.81 $\pm$ 3.33 \\
         \siii \SI{18.7}{\micro\meter} & & & & & & & 7.37 $\pm$ 1.25 \\
          \oiii  \SI{88}{\micro\meter}  & & & & & & & 33.70 $\pm$ 3.33 \\
          \cii \SI{157}{\micro\meter}  &  & & & & & & 3.14 $\pm$ 0.81 \\ 
    \bottomrule
    \end{tabular}
    }
  \\Notes: Fluxes are in 10$^{-18}$ Wm$^{-2}$.
  \label{tab:IR}  
\end{table}

\begin{table}

\caption{Emission line redshift determinations}
\resizebox{\columnwidth}{!}{%
\begin{tabular}{@{}lccr@{}}
\toprule
Emission lines & $\lambda_{Rest} (\si{\angstrom}) $ & $\lambda_{Obs} (\si{\angstrom})$ & z \\
 \midrule
O \textsc{iii}]$\rm{\lambda 1660}$ & 1660.81 & 1667.6 & 0.004086 \\
O \textsc{iii}]$\rm{\lambda 1666}$ & 1666.15 & 1672.95 & 0.004079 \\
C \textsc{iii}]$\rm{\lambda 1907}$ & 1906.68 & 1914.43 & 0.004066 \\
C \textsc{iii}]$\rm{\lambda 1909}$ & 1908.73 & 1916.46 & 0.004051 \\
\midrule 
z (Mean $\pm$ Std)$^a$  &&& 0.0040705 $\pm$ 0.000013\\
\bottomrule
\end{tabular}%
}
Notes: Mean and Std denote the mean and standard deviations of the redshifts estimated from the emission lines. 
\label{tab:z}
\end{table}

\section{Observations}
\label{section:observations}
\indent 

\subsection{HST/UV spectroscopy}
\label{section:hst}
The HST/COS observations were taken as part of the General Observing programs in HST Cycles 27 and 28 (GO: 16071 and  16445, PI: N Kumari) at lifetime adjustment position 4 (LP=4). Before taking each UV spectrum, the NUV target acquisition image is taken using the Mirror A. The FUV and NUV spectra were taken with the 2.5 arcsec diameter Primary Science Aperture (PSA) using the medium resolution gratings, G130M, G160M and G185M centred at 1291\AA, 1623\AA~ and 1913\AA, respectively. We used all FP-POS positions for better spectral sampling and increased signal-to-noise (S/N). Table \ref{tab:data} lists the exposure times for gratings and Mirror A used within the two HST programs. All HST/COS data were processed with the standard data reduction pipeline CALCOS version 3.4.0. 

\indent Figure \ref{fig:NUV/Acq} shows the HST/COS NUV target acquisition image where the red circle denotes the 2.5 arcsec COS spectroscopic aperture. The wavelength settings allow us to cover several spectral features consisting of ISM (red), photospheric (purple), wind (yellow) and nebular (brown) lines as shown in Figures \ref{fig:uv_spectra1} and \ref{fig:uv_spectra2}.

\begin{subfigures}
\begin{figure*}
    \centering
    \includegraphics[width=0.9\textwidth, trim={2cm 0.6cm 4cm 1.5cm}, clip]{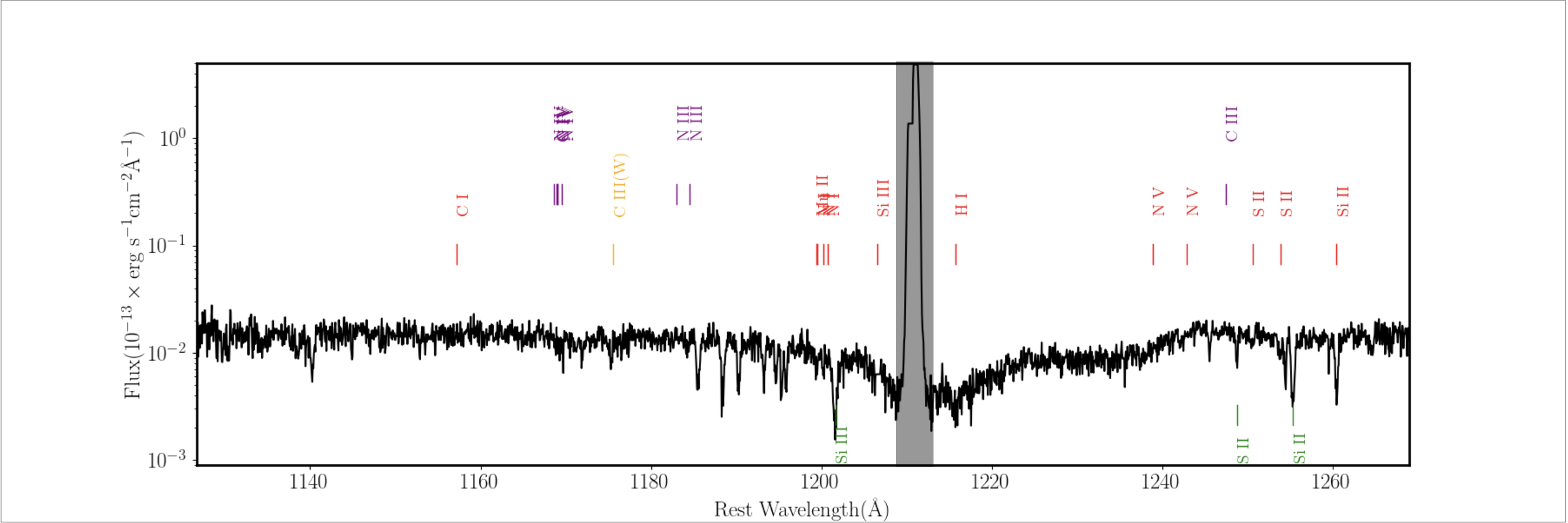}
    \includegraphics[width=0.9\textwidth, trim={2cm 0.6cm 4cm 1.5cm}, clip]{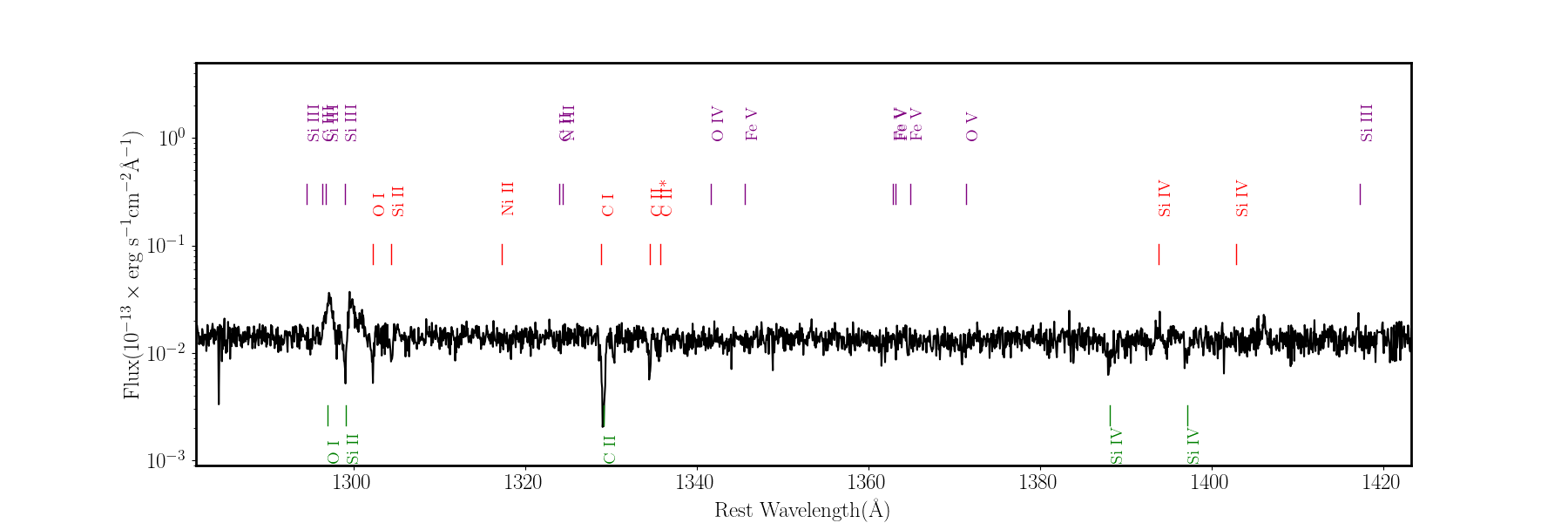}
    \includegraphics[width=0.9\textwidth, trim={2cm 0.6cm 4cm 1.5cm}, clip]{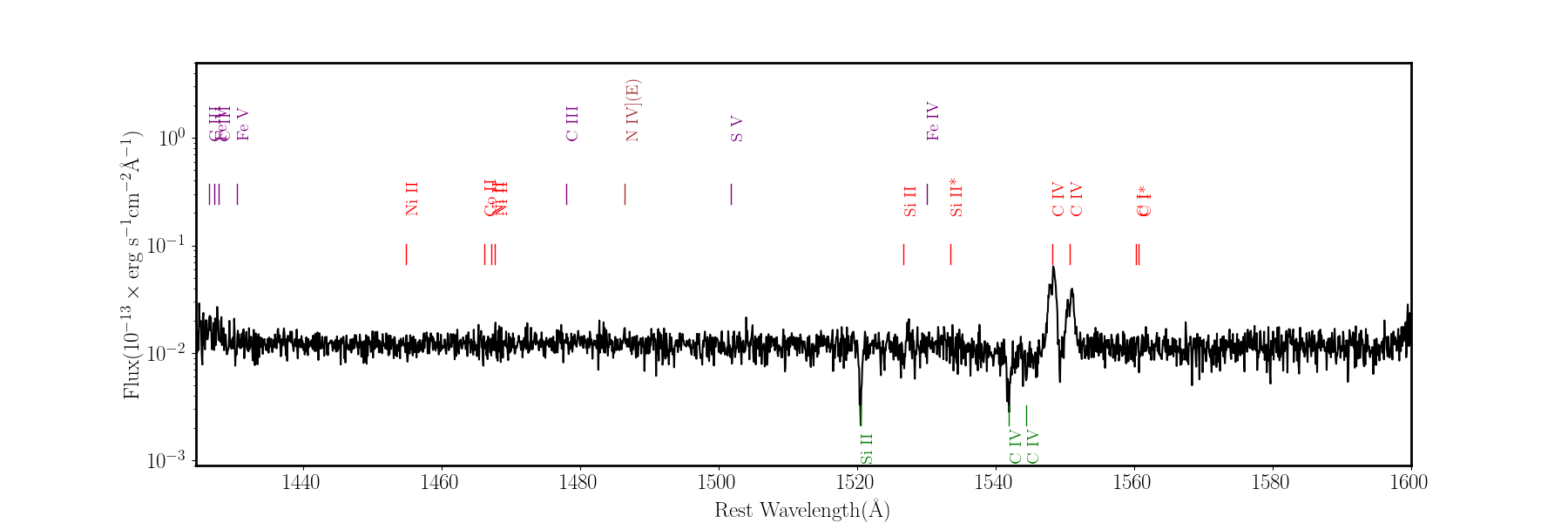}
    \includegraphics[width=0.9\textwidth, trim={2cm 0cm 4cm 1.5cm}, clip]{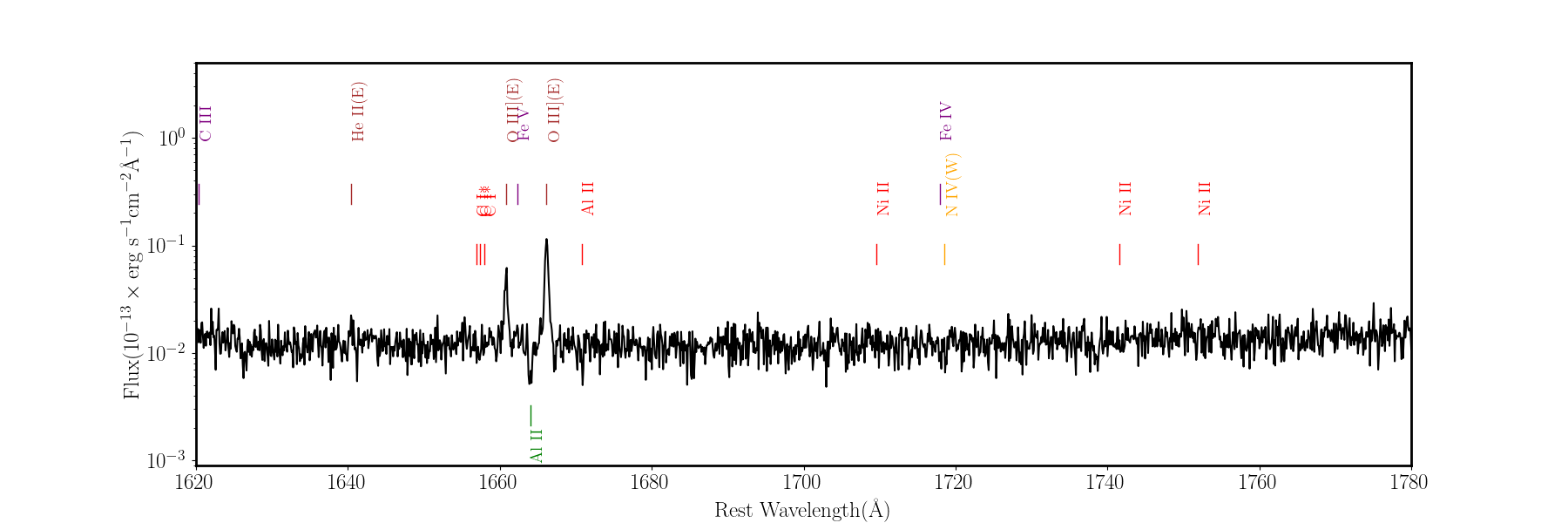}
       \caption{HST/COS FUV spectra of Pox 186 taken with G130M/1291 (upper two panels) and G160M/1623 (lower two panels). These fluxes are smoothed by a boxcar filter of 6 pixels for better visualization. These spectra show several spectral features consisting of ISM (red), photospheric (purple), wind (yellow) and nebular (brown) lines. The milky way lines are marked in green. Note that not all marked lines are detected, and are provided here for reference. The geocoronal emission is marked by a grey shaded region in the first panel.}
    \label{fig:uv_spectra1}
\end{figure*}

 \begin{figure}
    \centering

  \includegraphics[width=0.45\textwidth]{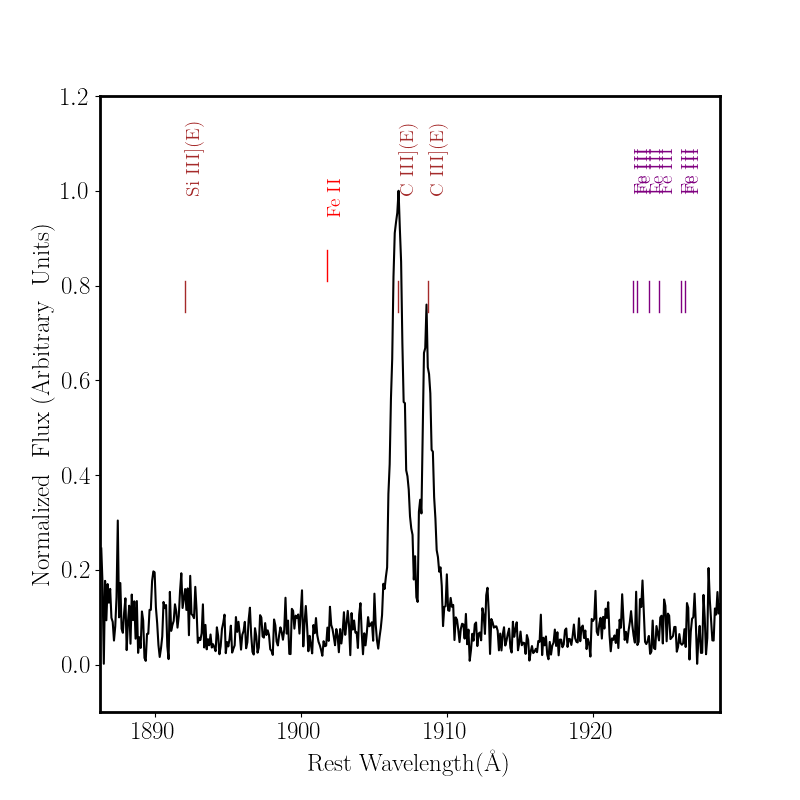}
  
    \caption{HST/COS FUV spectra of Pox 186 taken with G185M/1913. These fluxes are smoothed by a boxcar filter of 2 pixels for better visualization. Note that the shown spectrum only represents the middle segment of the G185M/1913 grating, and the other two segments do not show any spectral feature.}
    \label{fig:uv_spectra2}
\end{figure}
\end{subfigures}

\begin{figure*}
   \centering   
   \includegraphics[width=0.9\textwidth, trim={2cm 0.1cm 4cm 0.5cm}, clip]{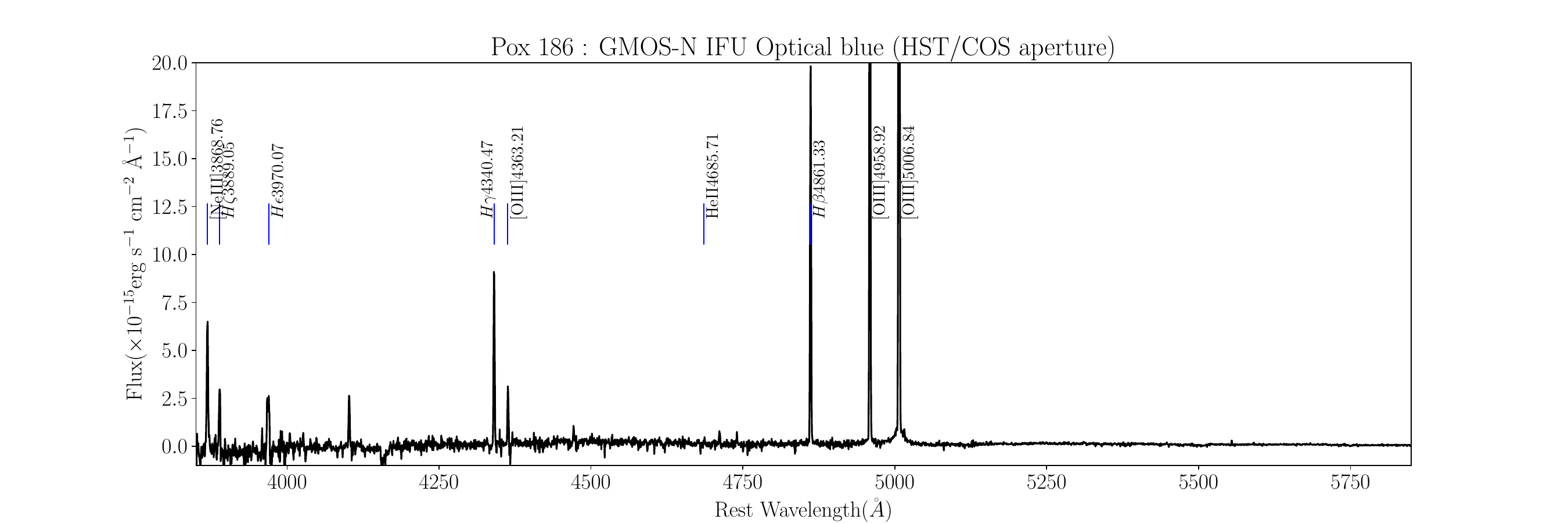}
    \includegraphics[width=0.9\textwidth, trim={2cm 0.1cm 4cm 0.5cm}, clip]{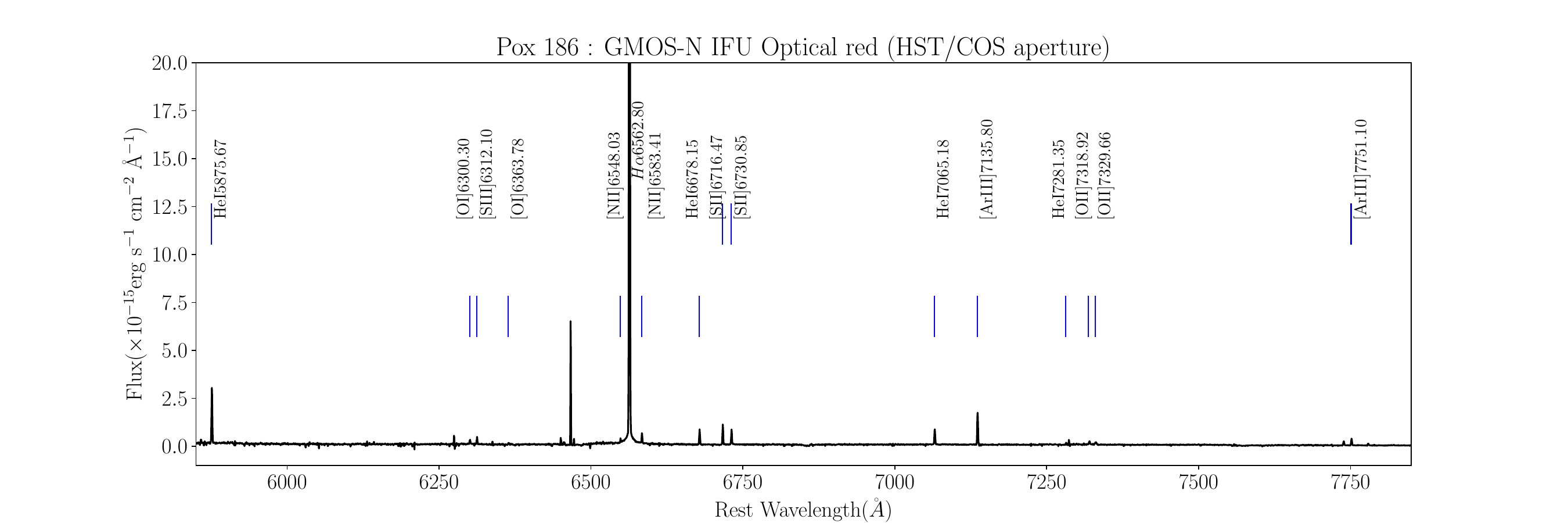}
    \includegraphics[width=0.9\textwidth, trim={2cm 0.1cm 4cm 0.5cm}, clip]{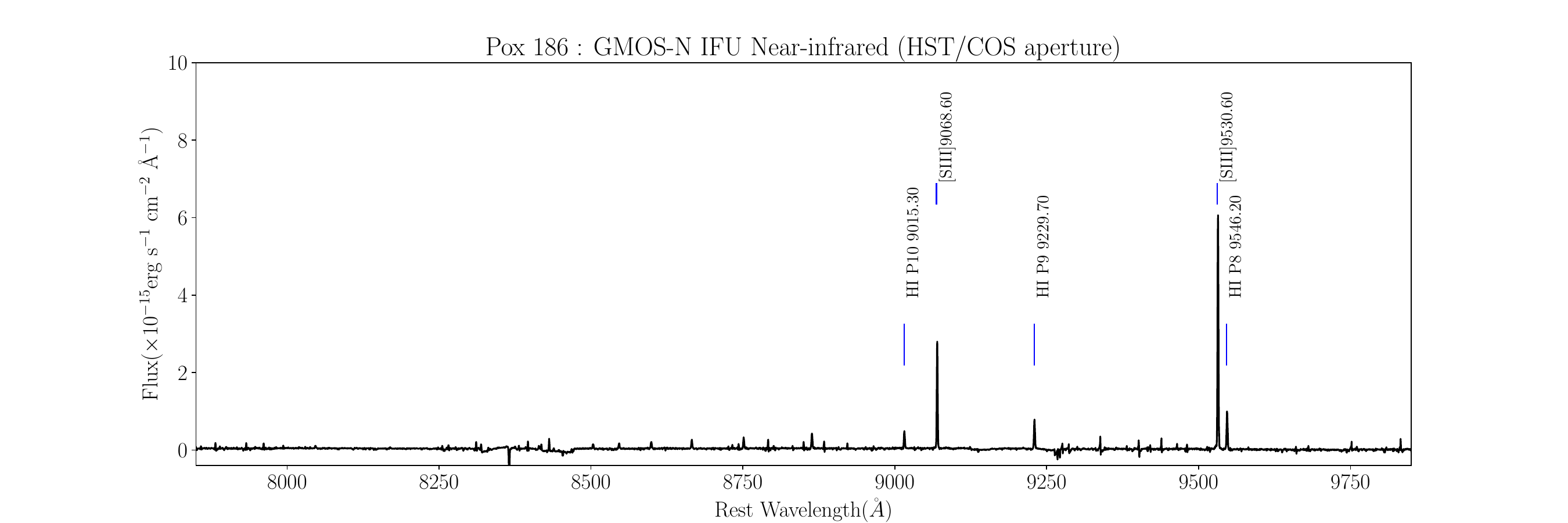}
    \caption{GMOS-N IFU integrated spectra of Pox 186 obtained by summing the spatially-resolved spectra within a circular aperture of radius 1.25$\arcsec$ overlapping with the COS aperture covering a rest-frame wavelength range of $\sim$ 3800--9900\AA. The important optical emission lines are marked in blue. We have also marked the location of He II 4686, which remains undetected in the optical spectrum.}
    \label{fig:optical spectra}
\end{figure*}

\subsection{GMOS-N optical spectroscopy}
\label{section:gmos}

\indent We obtained the spatially-resolved optical spectroscopy of Pox 186 using the GMOS \citep{Hook2004} and IFU \citep[GMOS-N IFU;][]{ Allington-Smith2002} at Gemini-North telescope in Hawaii, as part of two separate programs (PID: GN-2020A-FT-105, GN-2021A-FT-111, PI: N Kumari). The first program focussed on covering the optical wavelength range of $\sim$3500-- 8000\AA, while the second program allowed us to cover the near-infrared (NIR) wavelength range of $\sim$8000-10000\AA. The observations were taken in one-slit queue mode providing FOV of 3.5$\arcsec\times$5$\arcsec$, large enough to cover the entire galaxy  (Figure \ref{fig:fov}). Along with the science exposures, standard calibration observations were obtained including GCAL flats, CuAr  lamp for wavelength calibration and standard star HZ44 for flux calibration.

\indent We performed the basic steps of data reduction using the standard GEMINI reduction pipeline (version v1.15) written in Image Reduction and Analysis Facility (\textsc{iraf}, version v2.16)\footnote{\textsc{iraf} is distributed by the National Optical Astronomy Observatory, which is operated by the Association of Universities for Research in Astronomy (AURA) under a cooperative agreement with the National Science Foundation.}. These basic steps included bias subtraction, flat field correction, wavelength calibration, sky subtraction, and differential atmospheric correction finally producing the 3D data cubes, and have been described in detail in \citet{KumariPhDT}. New GMOS-N detectors were installed in 2017, which further required the quantum efficiency correction for all flats. We also used the L.A.Cosmic \citep{vanDokkum2001} to remove the cosmic rays from the science exposures. We chose a spatial sampling of 0.25$\arcsec$ for the final three-dimensional data cubes. We thus obtain three three-dimensional data cubes. We scale the flux of each of the three cubes using the methodology described in Appendix \ref{app:fluxscale}.

\subsection{Ancillary Mid-Infrared and Far-Infrared spectroscopy}
\label{sec:IR}
\indent \textit{MIR:} Pox 186 was observed with the Spitzer telescope in the low-resolution (R $\approx$ 60-127) mode using long-slits of the InfraRed Spectrograph in the \citep[IRS;][]{Houck2004}. The width of long slit is $\sim$ 3.6\arcsec, and its point-spread function (PSF) may extend beyond the slit width. \citet{Cormier2015} reports the MIR emission line fluxes after taking into account PSF.
\indent 

\noindent \textit{FIR:} Pox 186 was also observed with Herschel using the PACS which covers a total FOV of 47\arcsec$\times$47\arcsec and has a beam size of 9\arcsec and 12\arcsec at \SI{60}{\micro\meter} and \SI{160}{\micro\meter}, respectively. \citet{Cormier2015} finds that the Pox 186 is well-centred on the brightest spaxel of their FIR maps and hence reports FIR emission line fluxes by applying a point-source correction to the brightest spaxel.

\indent For completeness, in Table \ref{tab:IR} we tabulate the line fluxes of Pox 186 in the MIR and FIR wavelength regime taken from \citet{Cormier2015}

\section{Results}
\label{section:results}

\subsection{Source Redshift and Distance}
\label{sec:z}
\indent We determine the source redshift by measuring the observed wavelength of four strong UV emission lines \oiiiuv~ and \ciii~ as shown in Table \ref{tab:z}. We do not use the strong \civ ~emission line because it is double-peaked (Section \ref{sec:lineprofiles}). We do not use the optical emission lines because the zero-point of the wavelength calibration of GMOS-N data is not very well-constrained \citep{KumariPhDT}. The redshift of Pox 186 is 0.0040705$\pm$0.000013 from the presented HST/COS data and agrees with that of \citet{Guseva2004} and \citet{Eggen2021} within uncertainties. At H$_o$ = 70 \kms and $\Omega_{m}$ = 0.3,  the observed redshift corresponds to a  luminosity distance of 17.5 Mpc, and an angular scale of 84.1 pc per arcsec. However, we derive a value of 12.6 Mpc if we use Cosmicflows-3 \citep[][]{Kourkchi2020}.


\subsection{Line fluxes and equivalent widths}
\label{sec:line}

\subsubsection{UV}
\label{sec:UV line}
\indent Table \ref{tab:UV_flux_EW} presents the fluxes and equivalent widths (EW) of the UV emission lines \oiiiuv, \ciii~ and \civ. We estimate fluxes by summing the fluxes in the spectral line after subtracting a local linear continuum fitted to either side of the three doublets. The continuum level at the central wavelength of each emission line is used to estimate their equivalent widths. We correct the UV emission line fluxes using the E(B-V) value determined from optical Balmer decrement which is described later in Section \ref{sec:redenning}.

\begin{table}

\caption{Equivalent widths and emission line fluxes (observed $F_{\lambda}$ and intrinsic $I_{\lambda}$) of the UV nebular emission lines.}
\resizebox{\columnwidth}{!}{%
\begin{tabular}{lccr}
\toprule
Emission lines & EW  (\si{\angstrom})   & F$_{\lambda}$ & I$_{\lambda}$\\

  \midrule

\civ$\rm{\lambda 1548}$ & $5.21 \pm 0.22 $ & $0.50 \pm 0.02 $ & $0.99 \pm 0.09 $\\
\civ$\rm{\lambda 1550}$ & $2.54 \pm 0.18 $ & $0.27 \pm 0.02 $ & $0.53 \pm 0.06 $\\
\heii$\rm{\lambda 1640}$ & -- & $<$0.08$^a$ & $<$0.16$^a$ \\
\oiiiuv$\rm{\lambda 1660}$ & $2.21 \pm 0.29 $ & $0.23 \pm 0.03 $&$0.43 \pm 0.06 $ \\
\oiiiuv$\rm{\lambda 1666}$ & $5.74 \pm 0.38 $ & $0.58 \pm 0.04 $& $1.08 \pm 0.11 $\\
\ciii$\rm{\lambda\lambda 1907, 1909}$ & $35.85 \pm 0.73 $ & $2.56 \pm 0.05 $& $4.4 \pm 0.3 $\\
\bottomrule
\end{tabular}%
}
Notes:
Fluxes are in units of  $\times$ 10$^{-14}$ erg s$^{-1}$ cm$^{-2}$.\\
 $^a$ 2$\sigma$ upper-limit
\label{tab:UV_flux_EW}
\end{table}

\begin{table*}
\caption{Optical emission line flux measurements (relative to \hb = 100)  obtained from the two sets of integrated Gemini/IFU spectra: (i) the one overlapping with the COS aperture (ii) the other from the entire Gemini FOV. Observed line fluxes ($F_{\lambda}$) are extinction-corrected using E(B-V) to calculate the intrinsic line fluxes ($I_{\lambda}$). }
\begin{tabular}{lccccc}
\toprule
Line & $\lambda_{air}$ & F$_{\lambda}$ (COS) & I$_{\lambda}$ (COS) & F$_{\lambda}$ (Gemini FOV) & I$_{\lambda}$ (Gemini FOV) \\
\midrule
$\rm H\gamma$ & 4340.47 & $45.18 \pm 1.27 $ & $46.59 \pm 2.28 $ & $42.54 \pm 3.06 $ & $44.40 \pm 2.90 $ \\
$\rm [OIII]$ & 4363.21 & $14.38 \pm 0.77 $ & $14.81 \pm 0.89 $ & $12.93 \pm 1.84 $ & $13.47 \pm 2.17 $ \\
$\rm H\beta$ & 4861.33 & $100.00 \pm 0.59 $ & $100.00 \pm 2.87 $ & $100.00 \pm 1.12 $ & $100.00 \pm 2.26 $ \\
$\rm [OIII]$ & 4958.92 & $212.80 \pm 1.50 $ & $211.71 \pm 7.74 $ & $216.59 \pm 2.82 $ & $215.04 \pm 8.85 $ \\
$\rm [OIII]$ & 5006.84 & $637.53 \pm 3.98 $ & $632.70 \pm 20.53 $ & $645.74 \pm 7.81 $ & $638.95 \pm 23.62 $ \\
$\rm  HeI$ & 5875.67 & $12.29 \pm 0.14 $ & $11.78 \pm 0.39 $ & $12.12 \pm 0.22 $ & $11.43 \pm 0.30 $ \\
$\rm [OI]$ & 6300.3 & $1.01 \pm 0.09 $ & $0.96 \pm 0.08 $ & $1.09 \pm 0.23 $ & $1.01 \pm 0.19 $ \\
$\rm [SIII]$ & 6312.1 & $1.46 \pm 0.11 $ & $1.38 \pm 0.12 $ & $1.49 \pm 0.27 $ & $1.38 \pm 0.26 $ \\
$\rm [NII]$ & 6548.03 & $0.71 \pm 0.15 $ & $0.67 \pm 0.11 $ & $0.80 \pm 0.16 $ & $0.74 \pm 0.13 $ \\
$\rm H\alpha$ & 6562.8 & $303.97 \pm 1.84 $ & $286.00 \pm 9.24 $ & $311.30 \pm 3.74 $ & $286.00 \pm 8.27 $ \\
$\rm [NII]$ & 6583.41 & $2.14 \pm 0.44 $ & $2.01 \pm 0.34 $ & $2.41 \pm 0.47 $ & $2.21 \pm 0.40 $ \\
$\rm HeI$ & 6678.152 & $2.78 \pm 0.06 $ & $2.61 \pm 0.10 $ & $2.67 \pm 0.13 $ & $2.44 \pm 0.11 $ \\
$\rm [SII]$ & 6716.47 & $3.67 \pm 0.07 $ & $3.44 \pm 0.11 $ & $3.99 \pm 0.16 $ & $3.64 \pm 0.17 $ \\
$\rm [SII]$ & 6730.85 & $2.81 \pm 0.07 $ & $2.64 \pm 0.09 $ & $3.15 \pm 0.16 $ & $2.88 \pm 0.19 $ \\
$\rm [ArIII]$ & 7135.8 & $6.81 \pm 0.09 $ & $6.32 \pm 0.20 $ & $7.20 \pm 0.20 $ & $6.49 \pm 0.22 $ \\
$\rm [OII]$ & 7318.92 & $0.97 \pm 0.08 $ & $0.89 \pm 0.07 $ & $18.47 \pm 0.54 $ & $16.55 \pm 0.63 $ \\
$\rm [OII]$ & 7329.66 & $0.85 \pm 0.07 $ & $0.79 \pm 0.05 $ & ... & ... \\
HI P10 & 9017.4 & $2.09 \pm 0.06 $ & $1.88 \pm 0.06 $ & --  & -- \\
$\rm [SIII]$ & 9068.6 & $11.45 \pm 0.11 $ & $10.25 \pm 0.12 $ & -- & -- \\
$\rm [SIII]$ & 9530.6 & $26.64 \pm 0.23 $ & $23.7 \pm 0.2 $ & -- & --\\

E(B-V) & & $0.059 \pm 0.007$  & & $0.083 \pm 0.009$ \\
F(\hb) & &  $40.76\pm  0.23$ & & $48.12 \pm 0.54$\\
\hline
\bottomrule
\end{tabular}
\\
Notes: F(\hb) in units of $\times$ 10$^{-15}$ erg cm$^{-2}$ s$^{-1}$.\\
...:  S/N $<$ 3 for a given line.\\
--: For these lines, the covered Gemini FOV is different than for the rest of the emission lines, hence we do not report their values. 
\label{tab:optical fluxes}
\end{table*}

\begin{table}
\caption{Summary of physical properties obtained from the Gemini optical integrated spectra and the COS UV spectra (value$\pm$uncertainty)}
\resizebox{0.9\columnwidth}{!}{%

\begin{tabular}{lcccr}
\toprule
Parameter & & & & Value \\
\midrule 

\oiii/\hb &&&& $0.805 \pm 0.003 $ \\

\nii/\ha &&&& $-2.15 \pm 0.09$ \\

\sii/\ha &&&& $-1.671 \pm 0.007$ \\
\\

EW(\oiii\lmlm4959,5007+\hb) (\AA) &&&& $1800 \pm 800$\\
EW(\ha) (\AA) &&&& $820 \pm 20$\\
$\beta$-slope &&&& $-0.36 \pm 0.02$\\

\\
T$\rm_e$([OIII]) ($\times$ 10$^4$ K)  && & &1.63 $\pm$ 0.05 \\
T$\rm_e$([OII]) ($\times$ 10$^4$ K)  && & & 1.39 $\pm$ 0.08 \\
N$\rm_e$ ([SII]) (cm$^{-3}$) &&&& 110 $\pm$ 30 \\

\\
12 + log(O$^+$/H$^+$) & & &&7.22 $\pm$ 0.12 \\
12 + log(O$^{2+}$/H$^+$) & & &&7.76 $\pm$ 0.04 \\
12 + log(O/H) & & &&7.87 $\pm$ 0.04 \\

\\
C$^{2+}$/O$^{2+}$ && &&$0.21 \pm 0.03 $ \\
C$^{3+}$/O$^{2+}$ && &&$0.041 \pm 0.004 $ \\
log(C/O)$\rm^{direct}$ &&&& $-0.59 \pm 0.04 $ \\
log(C/O)$\rm^{empirical}$ &&&& $-0.60 \pm 0.03 $ \\

\\


T$\rm _{eff} \ddagger$ (kK) &&&& $60 \pm 18$ \\
log $\mathcal{U} \ddagger$ &&&& $-2.4 \pm 0.4$ \\
log (q/cm s$^{-1} \ddagger$) &&&& $8.1 \pm 0.4$\\
\\
log $\mathcal{U} \star$ &&&& $-2.661 \pm 0.014$\\

\bottomrule
\end{tabular}%
}

Notes: 
$\ddagger$: Output from \textsc{hcm-teff} assuming spherical geometry.
$\star$: Ionization parameter using \siii/\sii calibration from \citet{Kewley2019}. 
\label{tab:abun}
\end{table}

\subsubsection{Optical}

\indent We want to compare the UV, optical, MIR and FIR properties of Pox186 together for which we need to take into account the varying aperture sizes or FOVs of the different instruments with which these four different datasets are acquired. The GMOS-IFU data allow us to probe the optical properties for different apertures and sizes. We chose to extract two sets of integrated spectra. The first one is obtained by integrating all GMOS spectra within a circular aperture of 1.25\arcsec radius centred on the brightest knot of Pox18, hence coinciding with the HST/COS aperture, and is referred to as `COS-matched integrated' spectra. While the second set of spectra is obtained by integrating all spectra within the GMOS-IFU FOV and is referred to as `Gemini-FOV integrated' spectra. The main difference between the two sets of spectra is that the COS-matched integrated spectra only include the compact core of Pox 186, and exclude its plume, unlike the Gemini-FOV integrated spectra which include both.  Figure \ref{fig:optical spectra} shows the COS-matched integrated spectra along with several optical and NIR emission lines. 

\indent We measure the emission line fluxes for the recombination and collisionally excited emission lines (except H$\alpha$ and \oiii\lm5007) within the integrated spectra by fitting single Gaussian profiles after subtracting a linear continuum in the spectral region of interest via custom-written python codes using \textsc{lmfit} package \citep{newville2014}. Equal weight is given to flux in each spectral pixel while fitting Gaussians and the fitting uncertainties on the Gaussian parameters are propagated to calculate the flux uncertainty. We also create emission line flux maps for all lines (including H$\alpha$ and \oiii\lm5007) of the entire GMOS FOVs in the same way, which are shown in Appendix \ref{app:maps} (Figures \ref{fig:a_flux} and \ref{fig:a_flux_nir}).

\indent The strong emission lines \ha and \oiii\lm5007 show a weak broad component in the integrated spectra matching COS aperture. We estimate the  \ha line flux by summing the fluxes under the line within the continuum subtracted spectrum and removing the flux contribution from the \nii lines  in the wings of \ha broad component. The wing of \oiii\lm5007 also shows the He I 5015 line, which we remove by modelling it with a Gaussian from the continuum subtracted spectrum. This spectrum is then used to estimate \oiii\lm5007 line flux by summing over the fluxes under the line. Tables \ref{tab:optical fluxes} presents the measured observed line fluxes for the emission lines measured in the two sets of the integrated GMOS spectra, the COS-matched integrated spectra (F$_{\lambda}$(COS)) and the Gemini-FOV integrates spectra (F${\lambda}$ (Gemini FOV)). The uncertainties on the emission line fluxes presented in the Table are the random measurement uncertainties. However, we also include a systematic flux uncertainty of 50\% (see Appendix \ref{app:fluxscale}) and propagate in the inferred properties whenever it becomes relevant in the  analysis, and is explicitly mentioned in the paper.

\subsection{Reddening correction}
\label{sec:redenning}
\indent We estimate the colour excess E(B-V), by using the attenuation curve of the Small Magellanic Cloud \citep[SMC;][]{Gordon2003} along with the observed Balmer decrement (\ha/\hb) assuming a Case B recombination and an electron temperature and density of 10$^4$K and 100 cm$^{-3}$, respectively. We estimate E(B-V) for both sets of integrated optical spectra, the one overlapping with COS and the other one corresponding to the entire GMOS FOV. The E(B-V) for the COS-matched integrated spectra is 0.053 $\pm$ 0.004 which is close to the E(B-V) of the Milky Way in the line-of-sight of Pox 186, i.e., 0.0385 $\pm$ 0.0016 \citep{Schlafly2011}, hence indicating a very low amount of dust in the central region of Pox 186.   

\indent The intrinsic fluxes for the UV and optical lines are estimated by correcting the observed line fluxes using the E(B-V). No reddening correction is done for the MIR or FIR emission line fluxes. Table \ref{tab:UV_flux_EW} presents the intrinsic fluxes of the emission lines of the UV COS spectra, while Table \ref{tab:optical fluxes} shows the intrinsic fluxes for the COS-matched and Gemini-FOV integrated spectra. The uncertainties on the intrinsic fluxes are derived from propagating the random uncertainties on fluxes measured while fitting the emission lines.

\subsection{Physical properties of ionized gas and ionizing stellar population}
\label{sec:physical}

\begin{figure}
    \centering
    \includegraphics[width=0.9\columnwidth]{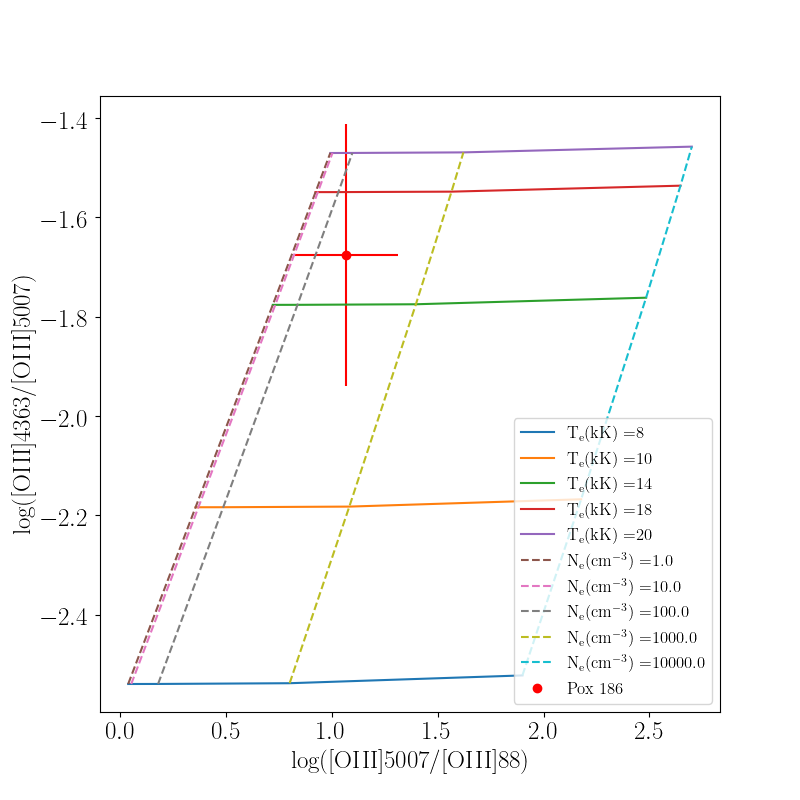}
    \caption{\textsc{pyneb} grids of \oiii\lm4363/\oiii\lm5007 versus \oiii\lm5007/\oiii\SI{88}{\micro\meter} for T$_e$=8--20kK and N$_e$=1--10000 cm$^{-3}$. The red point denotes the emission line flux ratios of Pox 186, where the optical \oiii~ emission line fluxes correspond to the entire Gemini-FOV, and the FIR \oiii~ line is taken from \citet{Cormier2015}, hence both optical and FIR datasets shown here cover the entire galaxy. The uncertainties on the line ratio include the systematic uncertainties on the optical line fluxes (see Section \ref{app:fluxscale}).}
    \label{fig:O3IR_optical}
\end{figure}

\subsubsection{Electron temperature and Density}
\label{sec:neb}
The UV, optical and IR spectra have emission lines sensitive to the electron temperature (T$_e$) and density (N$_e$). For example, the UV line \oiiiuv \lmlm 1660,1666 is temperature-sensitive and when combined with the optical \oiii \lm5007 line, probes T$_e$. Similarly, the optical line ratio of \oiii \lm 4363 and \oiii \lmlm 4959, 5007 is also sensitive to T$_e$. Similarly, the optical to IR line ratio, \oiii \lm 5007/\oiii \lm \SI{88}{\micro\meter} is sensitive to T$_e$, but also to N$_e$ \citep{Dinerstein1985} . 

\indent Figure \ref{fig:O3IR_optical} shows the optical \oiii~ lines flux ratio  (\oiii \lm 4363 / \oiii \lm 5007) versus the optical-IR \oiii~ lines flux ratio \oiii \lm 5007/\oiii \lm \SI{88}{\micro\meter}, where the grids are generated using the emissivities of the respective \oiii ~lines from  the \textsc{pyneb} code for a set of T$_e$ and N$_e$ values. The grid lines are not orthogonal as  \oiii \lm 5007/\oiii \lm \SI{88}{\micro\meter} is sensitive to both T$_e$ and N$_e$. We also show the optical \oiii~ lines flux ratio and optical-IR \oiii~ lines flux ratio \oiii of Pox 186, obtained from Gemini-FOV integrated spectra. We deem the use of Gemini-FOV integrated spectra for this comparison instead of the COS-matched integrated spectra because of the large FOV and PSF of the Herschel data (see Section \ref{sec:IR}) which includes not only the compact core of Pox 186 but also its plume. Note that the COS-matched integrated spectra miss the plume of Pox 186. The observed lines ratio lies on the \textsc{pyneb} grids and is in contradiction to that found by \citet{Chen2023}.  The optical \oiii~ emission line flux ratio for the Gemini-FOV integrated spectra corresponds to T$_e$ = 15000 $\pm$ 1300K. 

\indent From the COS-matched optical spectra, we estimate T$_e$ (\oiii) =  16300$\pm$500K  by using the emission line flux ratio of the auroral line \oiii \lm 4363 and \oiii \lmlm 4959, 5007. Since UV line O \textsc{iii}] $\lambda\lambda$ 1660, 1666 is also temperature-sensitive, we also estimate T$_e$(\oiii) by using the emission line flux ratio of \oiii\lm5007/\oiiiuv\lmlm1660,1666, which is in agreement with that obtained from optical emission line ratios. 
The electron temperatures derived from the COS-matched and Gemini-FOV integrated spectra agree with each other and are typical of the \hii regions within the star-forming galaxies \citep{Kumari2017, Kumari2018, Kumari2019}.

\indent We measure electron density using the density-sensitive line ratio \sii \lmlm 6717, 6731 line ratio and T$_e$ (\oiii) determined above for the COS-matched as well as the Gemini-FOV integrated spectra. For both datasets, the electron density indicates a low-density regime. We note that the density-sensitive line doublets \oii\lmlm 3727, 3729 could not be used for determining density, as the sensitivity of GMOS-IFU in the blue end has degraded over time, and hence the blue-end data are unusable. We do not estimate N$_e$ from the density-sensitive UV doublet C \textsc{iii}]  $\lambda\lambda$1907, 1909 available from the COS spectra as the doublet is blended and asymmetric.

\subsubsection{Chemical abundances}
\label{sec:metallicity}

 The chemical abundances of Pox 186 are only determined for the COS-overlapping central region, because of the non-detection of the necessary emission lines in the Gemini-FOV integrated spectra. 

\indent Gas-phase metallicity: The gas-phase metallicity (12+log(O/H)) can be robustly estimated from the T$_e$-base direct method, where the abundances of the dominant ionic states  of oxygen  (O$^+$ and O$^{2+}$) are first determined  from the temperatures of their respective ionization zones and then combined to estimate the total oxygen abundance.  The temperature of the high-ionization zone  T$_e$(\oiii) is determined as derived in Section \ref{sec:neb}  and is combined with N$_e$ (\sii) to estimate the temperature of the low-ionization zone by using the density-dependent calibration given in \citet{Perez-Montero2017}. Like \citet{Kumari2019} where \oii\lmlm3727, 3729 remain undetected, we measure  O$^+$/H$^+$ using \oii\lmlm 7320, 7330 and the low-ionization temperature T$_e$(\oii) by employing the formula given in \citet{Kniazev2003}. We measure O$^{2+}$/H$^+$ using \oiii \lmlm 4959, 5007 and T$_e$(\oiii) in the formula given in \citet{Perez-Montero2017}. The oxygen ionic abundances are combined to calculate the oxygen elemental abundance, 12 + log(O/H) = 7.87$\pm$0.04, for the region of Pox 186 probed by COS, and agrees within 3$\sigma$ with that derived by \citet{Guseva2004} for a larger region of this galaxy. 

\indent Carbon-to-oxygen ratio:  We follow the relations between T$_e$ (\oiii) and dereddened line ratios C \textsc{iii}]/O \textsc{iii}] and C \textsc{iv}/O \textsc{iii}] provided in \citet{Perez-Montero2017}, we estimate C$^{2+}$/O$^{2+}$ and C$^{3+}$/O$^{2+}$. We estimate direct method C/O by combining C$^{2+}$/O$^2+$ and C$^3+$/O$^2+$, assuming $\frac{C}{O} = \frac{C^{2+} + C^{3+}}{O^{2+}}$. We also estimate C/O using the empirical method given in \citep{Perez-Montero2017}. The estimates of C/O from the direct and empirical method are in excellent agreement with each other (Table \ref{tab:abun}). 


\subsubsection{Ionization parameter and Effective Temperature}
\label{sec:logUTeff}

For estimating the ionization parameter (log $\mathcal{U}$) and effective temperature (T$\rm_{eff}$) of the central region matching COS-aperture, we use the publicly available code \textsc{Hcm-Teff code} \citep[v5.3][]{Perez-Montero2019} where we assume a blackbody model and use reddening-corrected optical emission line fluxes (\oiii\lmlm4959, 5007, \sii\lmlm 6717, 6731,  He I\lm 6678, Ar \textsc{i}\lm 7135, \siii\lmlm 9069, 9532) from the COS-matched integrated and the gas-phase metallicity (12 + log(O/H) = 7.87$\pm$0.04, Section \ref{sec:metallicity}), and run the code for plane-parallel and spherical geometry separately. 
We find that the choice of geometry has no effect on either log $\mathcal{U}$ or  T$\rm_{eff}$ for the central region of Pox186. We also estimate log $\mathcal{U}$ = -2.661 $\pm$ 0.014 using the calibration involving \siii\lmlm9069,9532/\sii6717,6731 given by \citet{Kewley2019}, which agrees with that derived from \textsc{HCm-teff}.  

For determining log $\mathcal{U}$ for the entire Pox186 galaxy, we use calibrations from \citet{Kewley2019} including the MIR line ratios (Table \ref{tab:IR}), \neiii/\neii  which gives log $\mathcal{U}$ = -2.55 $\pm$ 0.22.  Note that we do not use optical line ratio \siii/\sii mainly because \siii \lmlm 9069, 9530 lines cover a slightly different FOV than the rest of the optical emission lines including \sii.  It is for the same reason that we could not use \textsc{HCm-teff} to estimate these two parameters for the Gemini FOV.

\subsubsection{Radiation hardness}
\label{sec:neta}

Hardness of the radiation field can be measured by the softness parameter log $\eta$. It was initially defined in terms of optical ionic ratios $\rm \eta =  (O^+/O^{2+})/(S^+/S^{2+})$ \citep{Vilchez1988}, and can be estimated from the calibration provided in \citet{Kumari2021}, i.e.,  $\rm log~ \eta~ =~ log~ \eta\prime~ +~ 0.16/t +~ 0.22 $, where t = T$_e$/10$^4$ and log $\eta\prime$ can be determined from the optical emission line ratios ( $\rm log~ \eta\prime = \frac{\oii/\oiii}{\sii/\siii}$) or the MIR infrared line ratios ( $\rm log~ \eta\prime = \frac{\neii/\neiii}{\siii/\siv}$) as mentioned in \citet{Perez-Montero2009a}. 

We estimate log $\eta\prime$ = -0.47 $\pm$ 0.23 using MIR emission line fluxes (Table \ref{tab:IR}). We estimate log $\eta$ = -0.14 $\pm$ 0.23 using T$_e$ from the Gemini-FOV integrated spectra and log $\eta\prime$ estimated earlier. We chose to use T$_e$ from the Gemini-FOV integrated spectra rather than COS-matched integrated spectra as the former covers the entire galaxy like MIR data. We do not use the optical emission line fluxes to determine log $\eta\prime$ as \oii\lmlm3729, 3729 are not detected because of the decreased sensitivity of GMOS-N IFU in the blue wavelength end. Lower log $\eta$ and log $\eta\prime$ indicate a hard radiation field.  Pox 186 exhibits lower values of log $\eta\prime$ and log $\eta$ compared to the average values exhibited by star-forming regions or galaxies in the local Universe \citep{Kumari2021}, thus indicating that the radiation field in Pox 186 is harder than average.

\indent 

\indent 

\indent Table \ref{tab:abun} summarizes all the physical properties derived in this section. The uncertainties on the derived quantities are estimated from the random uncertainties on the flux measured while fitting emission lines and excluding the systematic flux uncertainty.

\section{Discussion}
\label{section:discussion}
\subsection{Large equivalent widths of nebular carbon lines}
\label{sec:carbon}

\begin{figure}
    \centering
    \includegraphics[width=0.45\textwidth]{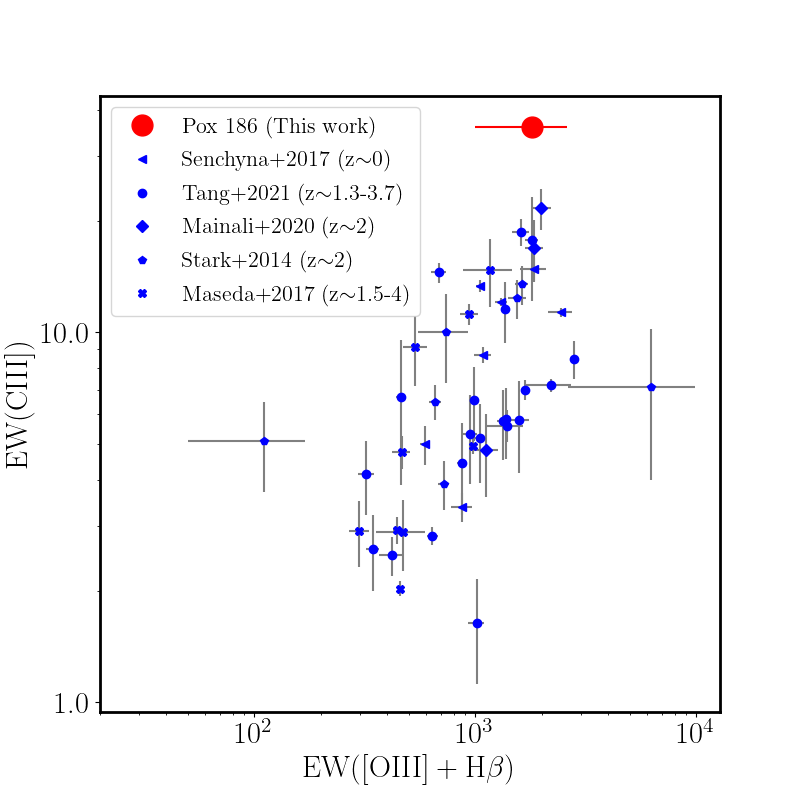}
    \caption{EW(\ciii) versus EW(\oiii + \hb) for Pox 186 (red dot) compared with galaxies at z$\sim$0-4, taken from \citet{Stark2014, Maseda2017, Senchyna2017, Mainali2020, Tang2021}}
    \label{fig:c3o3}
\end{figure}

\indent We find EW(\ciii\lmlm1906,1909) = 35.85 $\pm$ 0.73\AA ~for Pox 186. Figure \ref{fig:c3o3} shows a comparison of EW(\ciii\lmlm1906,1909) versus EW(\oiii\lmlm4959,5007+\hb) for Pox 186 with galaxies at z$\sim$0--4. We find that Pox 186 exhibits the highest EW(\ciii) among all the star-forming galaxies in the local Universe \citep[z $\sim$ 0,][Figure \ref{fig:c3o3}]{Leitherer2011, Senchyna2017, Senchyna2019}, and the majority of the star-forming galaxies at the intermediate \citep[z$\sim$ 2--4,][]{Erb2010, Stark2014, Vanzella2016, Maseda2017, Vanzella2017, Tang2021}. However, EW(\ciii) is also reported to lie within 20--40\AA~ for a small fraction ($\sim$1.2\%) of star-forming galaxies at the intermediate \citep[z$\sim$ 2--4,][]{Lefevre2019} and a few EoR galaxies \citep[z$\gtrsim$6,][]{Stark2017, Hutchison2019, Topping2021, Jiang2021}. 

\indent We also measure the EW(\civ\lmlm1548,1550) = 7.75$\pm$0.28\AA~ for Pox 186, which is comparable to those found for local metal-poor galaxies with very young stellar populations \citep{Berg2016, Berg2019, Senchyna2019}. 

\indent We explore the cause for the extreme EW of carbon lines in the following:

\subsubsection{High effective temperature}
\label{sec:highT} 
\indent \citet{Nakajima2018} states that  EW(\ciii) >30\AA~ can be caused by blackbody with extremely high effective temperature (T$\rm _{eff}$), i.e. > 6$\times$10$^4$K. We estimate T$\rm _{eff}$ = 60 $\pm$ 18 kK assuming blackbody using the \textsc{Hcm-teff} code (Section \ref{sec:logUTeff}), thus indicating that the high effective temperature may be responsible for high carbon EW observed for Pox 186. 

\begin{figure}
    \centering
    \includegraphics[width=\columnwidth]{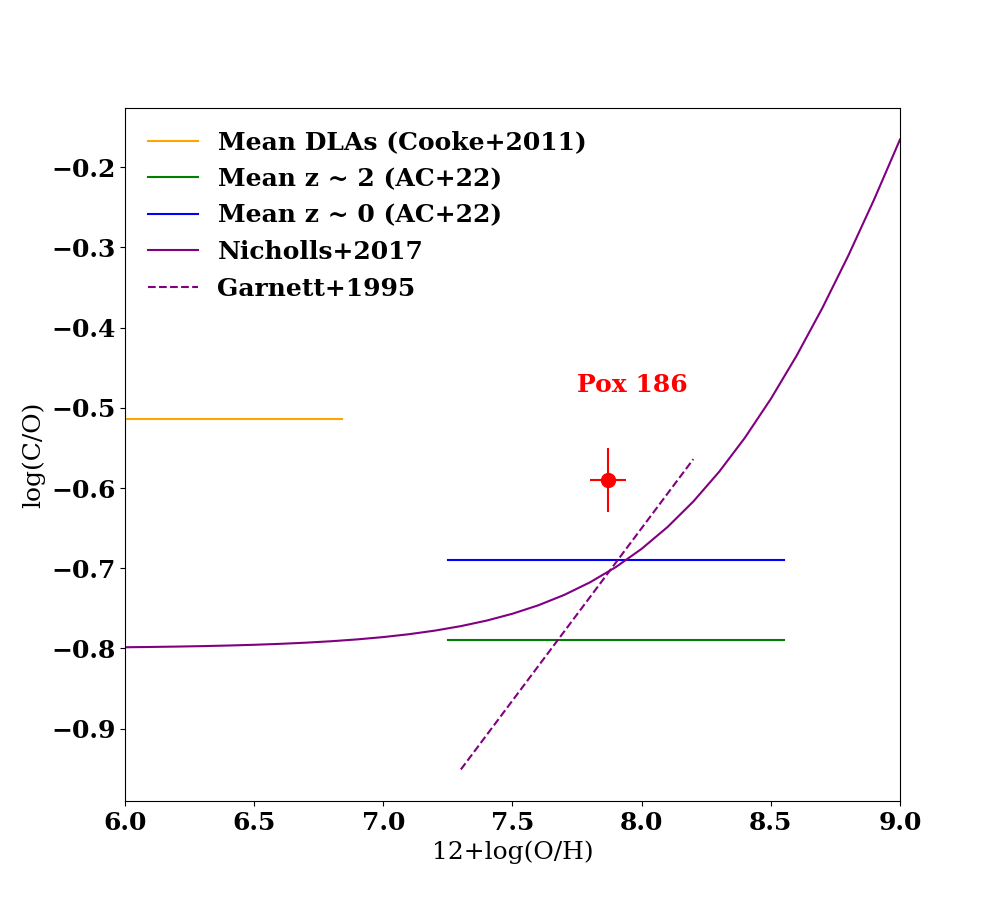}
    \caption{C/O versus O/H for Pox 186, along with average C/O for z$\sim$0 (horizontal blue line) and z$\sim$2 (horizontal green line) from \citet{Arellano-Cordova2022} and for DLAs from \citet{Cooke2011}. The solid and dashed purple lines indicate the C/O-O/H relation given in \citet{Nicholls2017} and \citet{Garnett1995}, respectively.}
    \label{fig:co_oh}
\end{figure}

\begin{figure*}
    \centering
    \includegraphics[width=0.9\textwidth]{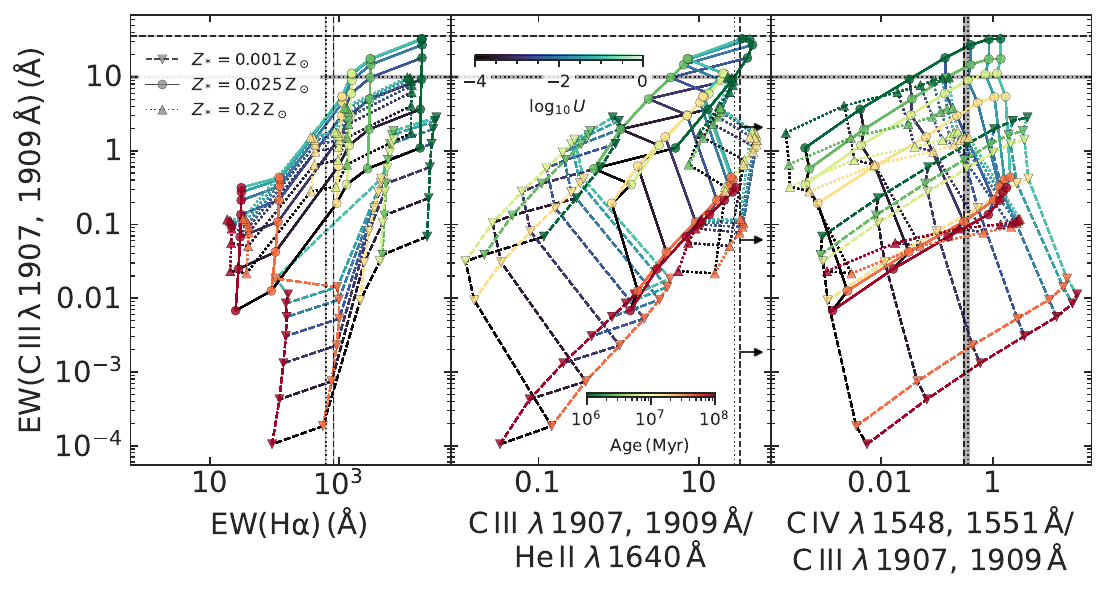}
    \caption{EW(\ciii) plotted against EW(\ha) (left-hand panel), \ciii/He II 1640 line ratio (middle panel) and \civ/\ciii ~(right-hand panel) using the \textsc{cloudy} models described in Section \ref{sec:cloudy}. The horizontal dashed line indicates the observed EW (\ciii)  while the horizontal dotted line indicates the reddening-corrected EW(\ciii) where line flux is corrected using nebular E(B-V) while continuum is corrected using the stellar E(B-V). Similarly, the vertical dashed and dotted lines indicate the observed and reddening-corrected quantities, respectively. The reddening-corrected EW is obtained by using nebular E(B-V) for line fluxes and stellar E(B-V) for continuum, while the reddening-corrected emission line ratios are obtained by using nebular E(B-V) for both emission lines in the ratio. The right-ward pointing arrows in the middle panel indicates the lower limit on the \ciii/\heii~ where a 2$\sigma$-upper limit on \heii~ line is considered.}
    \label{fig:Cloudy_models}
\end{figure*}

\subsubsection{High carbon-to-oxygen ratio}
\label{sec:abundance} 
\indent \citet{Jiang2021} suggests that the high EW(\ciii) and EW(\civ) could be due to a higher carbon abundance. 

We explore this via Figure \ref{fig:co_oh}, which shows that the C/O abundance of Pox 186 (red point) is higher than the average for galaxies found in the same metallicity range at  z $\sim$ 0 (horizontal blue line) and z$\sim$ 2 (horizontal green line) from \citet{Arellano-Cordova2022}. It is also higher than that predicted by the best-fit line to C/O versus O/H for measurements of stars derived by \citet[][solid purple curve]{Nicholls2017} and for measurements of irregular dwarf galaxies derived from \citet[][dashed solid line]{Garnett1995}. Thus, the higher C/O for Pox 186 supports the argument from \citet{Jiang2021} about higher carbon abundance causing the higher EW of carbon lines. 

\indent We note that the optical spectrum of Pox186 does not show any signature of the carbon-rich Wolf-Rayet (WR) stars either as red or blue WR bump which could lead to a direct enhancement of carbon. \citet{Schaerer1999b} lists Pox 186 as a WR galaxy on the basis of a broad \heii\lm4686 at 0.8$\sigma$ above background reported in \citet{Kunth1985}. The high-quality HST/COS and GMOS-IFU data allow us to exclude Pox 186 as a WR candidate. 

\begin{figure*}
\centering 
\includegraphics[width=0.45\textwidth]{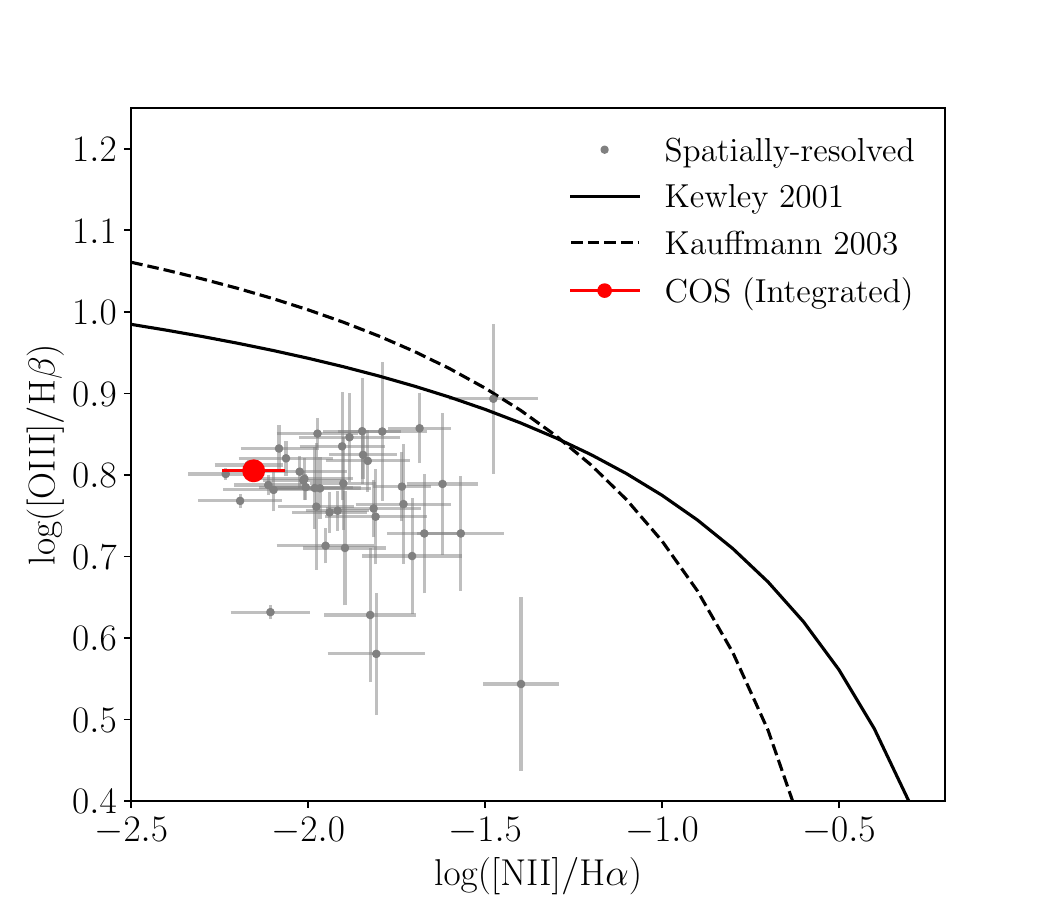}
\includegraphics[width=0.45\textwidth]{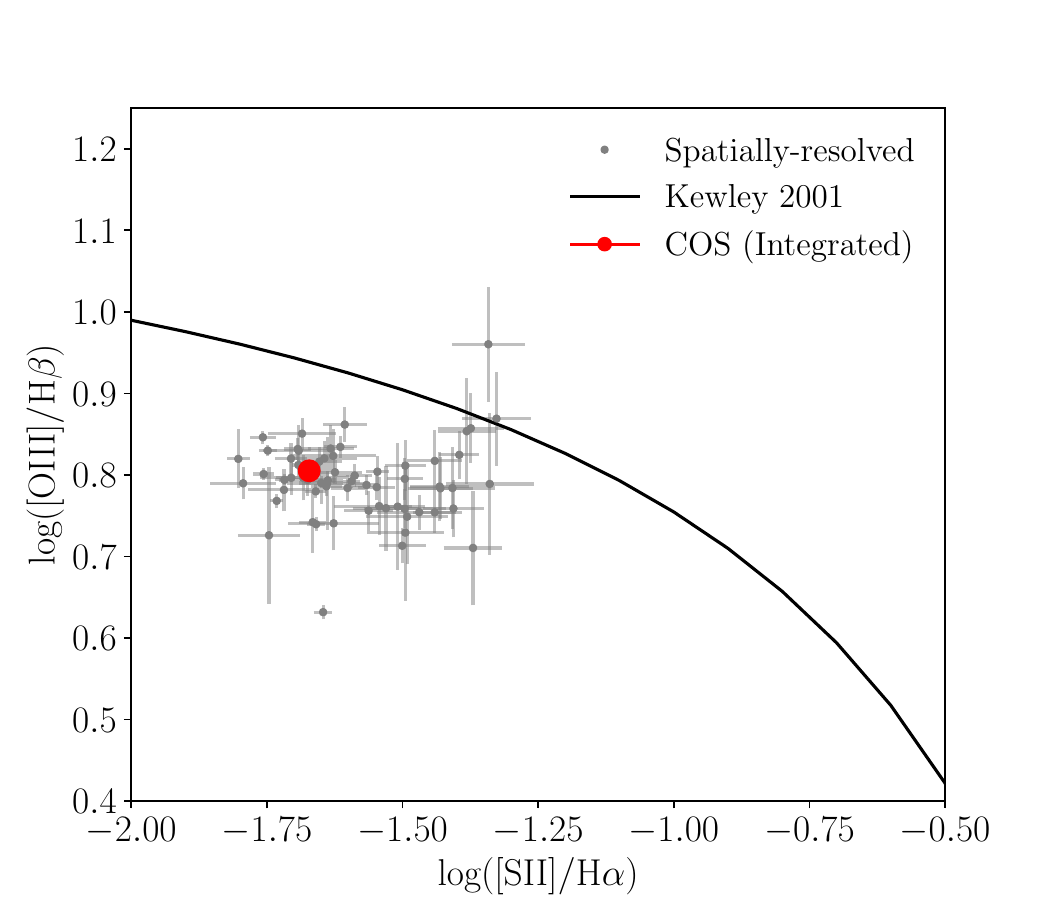}
\caption{Classical optical emission line ratio diagrams: \oiii/\hb versus \nii/\ha (left-hand panel) and \oiii/\hb versus \sii/\ha (right-hand  panel). Black solid curve and dashed curve represent the maximum starburst line from \citet[][theoretical]{Kewley2001} and \citet[][empirical]{Kauffmann2003}, respectively, distinguishing between the photoionized and non-photoionized regions. The blue dot denotes the corresponding emission line ratios obtained from the optical spectra integrated over the COS aperture, while the grey dots indicate the spatially-resolved emission line ratios.} 
\label{fig:bpt}
\end{figure*}

\subsubsection{Slope and upper mass of top-heavy initial mass function}
\label{sec:cloudy} 
To understand the origin of the extreme EW(\ciii) measured in Pox186, we consider \textsc{Cloudy} models broadly similar to those in \citet{Witstok2022}; here, we give a brief summary and highlight the differences with the models presented in \citet{Witstok2022}. The incident radiation field of a single burst of star formation with varying ages ($1 \, \mathrm{Myr}$ to $100 \, \mathrm{Myr}$) is generated by \textsc{bpass} v2.1 stellar population synthesis models including binary stars under a top-heavy initial mass function \citep[IMF; with slope $-2$;][]{Eldridge2017}, ranging in stellar mass from $1 \, \mathrm{M_\odot}$ to $300 \, \mathrm{M_\odot}$. Unlike in \citet{Witstok2022}, calculations stop when a molecular fraction of $10^{-6}$ is reached, such that the model does not extend into a photodissociation region beyond the central \hii\ region.

We considered models with a wide range of base (stellar) metallicities, tuning the gas-phase metallicity to match the observed values of Pox186. We introduce an additional nebular $\mathrm{\upalpha/Fe}$ enhancement, which is accomplished by increasing the nebular abundances of individual $\upalpha$ elements \citep[for details, we refer to][]{Witstok2022}. The nebular elemental abundances of the main $\upalpha$ elements (C, O, Ne, Mg, Si, S) are scaled up by $4 \times$, except for carbon which is increased by a factor of $2$, so that the $\mathrm{C/O}$ ratio is approximately half the solar value as appropriate for Pox186 (see Section \ref{sec:abundance}). Moreover, this implies our fiducial model with stellar metallicity $Z_* = 0.025 \, \mathrm{Z_\odot}$ has a nebular oxygen abundance of approximately $10\%$ solar, as directly measured for Pox186 (see Section \ref{sec:abundance}). We vary the ionisation parameter and hydrogen density between $-4 < \log_{10} U < -1$ and $10^{-1} \, \mathrm{cm^{-3}} < n_\mathrm{H} < 10^4 \, \mathrm{cm^{-3}}$ (as measured at the illuminated face of the cloud), respectively.
 
An overview of the models generated in \textsc{Cloudy} is shown in Figure \ref{fig:Cloudy_models}. For simplicity, we only show models with a fixed density of $n_\mathrm{H} = 10^2 \, \mathrm{cm^{-3}}$ and stellar metallicities of $0.001 \, \mathrm{Z_\odot}$, $0.025 \, \mathrm{Z_\odot}$, and $0.2 \, \mathrm{Z_\odot}$. We find particularly the slope and upper mass of the IMF are restrictive in reproducing the extreme EWs of \ciii, as models with an upper mass of $100 \, \mathrm{M_\odot}$ only reach EWs of approximately $20 \, \Angstrom$\footnote{For comparison, in Appendix \ref{app:bpass}, we present the models using an upper mass of $300 \, \mathrm{M_\odot}$ and slope -2.3 and no $\upalpha$/Fe enhancement which reproduce EW(\civ) and not the EW(\ciii).}. 
However, none of these models could simultaneously reproduce the observed EW of \ciii ~  and \ha emission lines, denoted by  dashed horizontal and vertical lines, respectively, in Figure \ref{fig:Cloudy_models}).

To explore this further, we considered the dust distribution which might affect the nebular \ciii ~and the underlying stellar continuum differently. We estimate the UV continuum slope $\beta$ = -0.36$\pm$0.02 using the spectral windows from \citet{Calzetti1994} in the wavelength region of $\sim$1250--1850, which indicates a red continuum. If we use this $\beta$ value to deredden the continuum using the SMC law from \citet{Reddy2018} and use E(B-V) derived from the optical data to deredden the emission line again using the SMC law, the EW(\ciii) may decrease by a factor of 3. The reddening-corrected values are shown by dotted vertical and horizontal lines, which can not be reproduced by the models either.

The inability of the models to reproduce the observed or reddening-corrected properties of Pox 186 could be due to the simplistic assumptions on the geometry and relative distribution of dust and gas within the photoionization models. In summary, it indicates the need to improve the existing population synthesis and photoionization models.

\begin{figure*}
    \centering
    \includegraphics[width=\columnwidth]{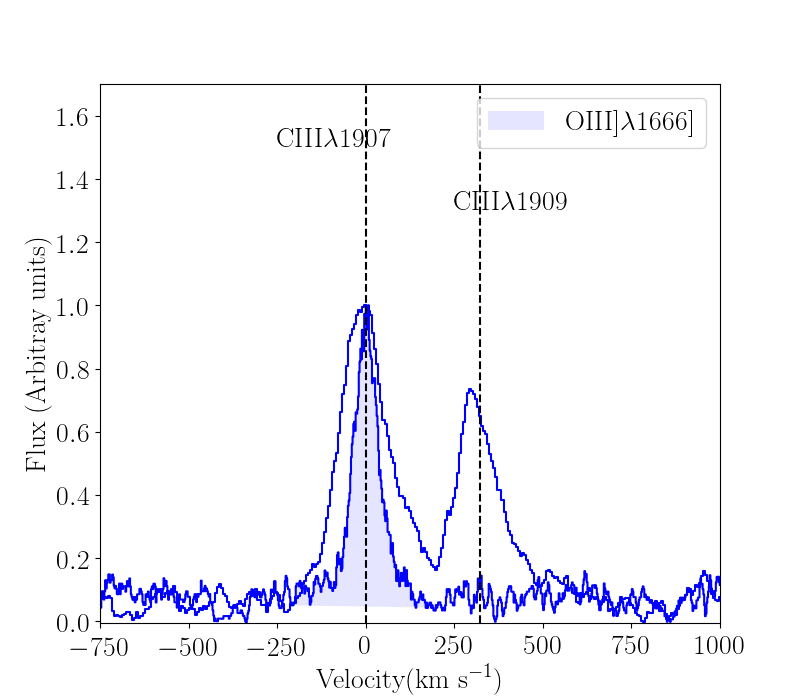}
     \includegraphics[width=\columnwidth]{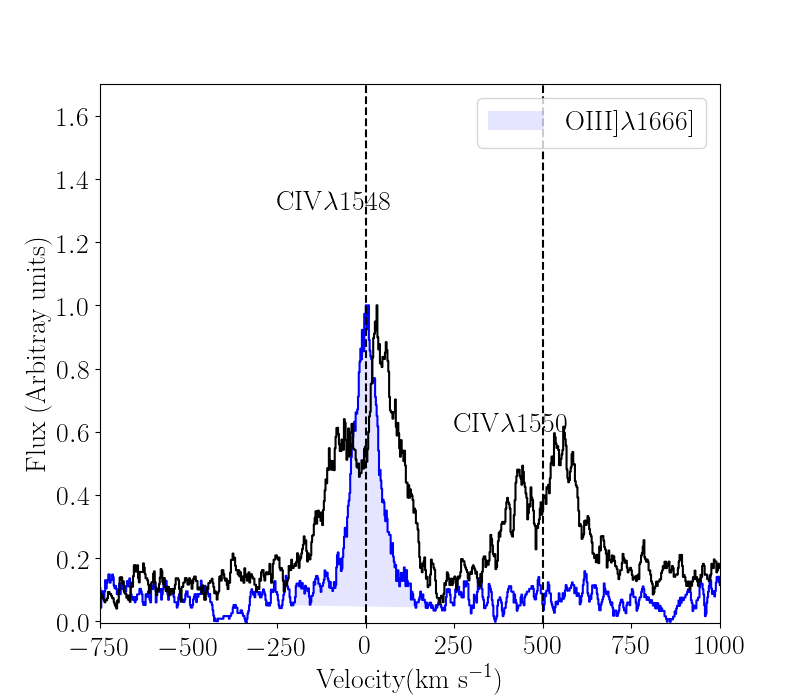}
    \caption{Comparison of \oiiiuv \lm 1666 line with respect to \ciii \lm 1907 (left-hand panel) and \civ \lm 1548 (right-hand panel). In both panels, all emission lines (\civ \lm 1548, \oiiiuv \lm 1666 and \ciii \lm 1907) are normalized by their peak fluxes.}
    \label{fig:overlapping}
\end{figure*}

\subsubsection{Hard Ionizing Radiation}
\label{sec:hardness}
\indent High EW(\ciii) is also proposed to be caused by the hard ionizing radiation from extreme stellar populations or AGN  \citep{Nakajima2018, Jiang2021}  or shocks from the radio jets \citep[][]{Best2000}. We rule out the possibility of an AGN/shocks causing the high EW(\ciii) as figure \ref{fig:bpt} shows that the optical emission line ratios obtained from the integrated spectrum, \oiii/\hb versus \nii/\ha (left-hand panel) and \oiii/\hb versus \sii/\ha (right-hand panel) do not occupy the AGN/shock region of the classical emission-line diagnostic diagrams \citep{Baldwin1981,  Veilleux1987}. Figure \ref{fig:bpt} also shows a few spaxel-based line ratios lying beyond the photoionization region on the BPT diagrams; however, they are too few ($\sim$2.5\% and $\sim$3\% for \nii-BPT and \sii-BPT diagrams, respectively) to be a conclusive indicator of hard AGN radiation. Hard ionizing radiation is also expected to produce \heii~ lines, so emission line ratio diagnostic diagrams including optical and UV \heii~ lines are also used to determine the presence of AGN \citep{Feltre2016, Brinchmann2008}. However, neither \heii~\lm 1640 nor \heii~\lm 4686 lines are detected in the spectrum corresponding to the COS pointing.  The above discussion shows that hard ionizing radiation from AGN is unlikely to be the cause of high EW(\ciii). 

\begin{figure}
    \centering
    \includegraphics[width=\columnwidth]{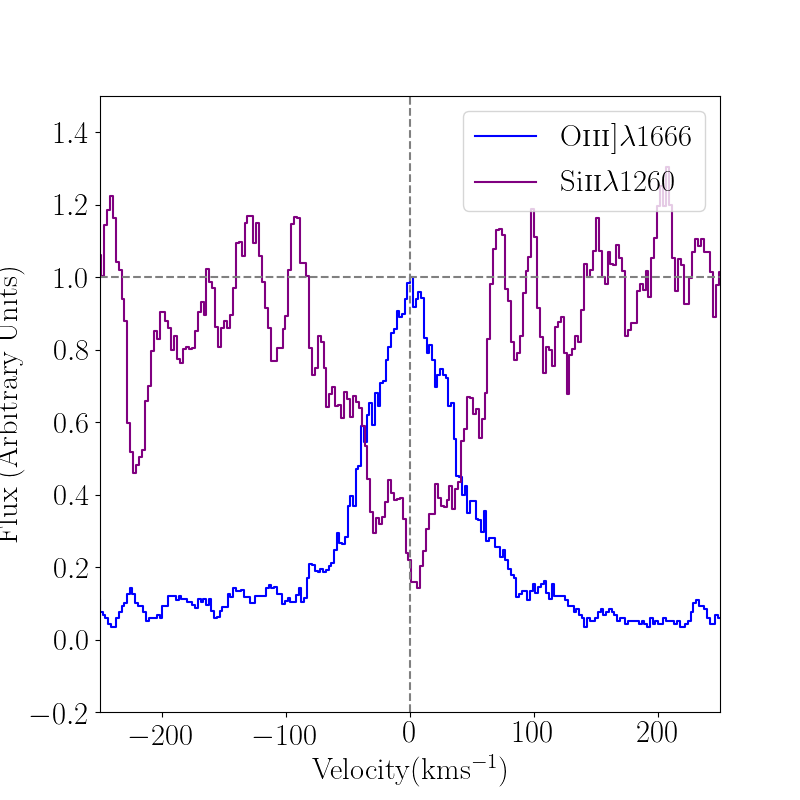}
    \caption{Comparison of velocity profiles of \oiiiuv \lm 1666 emission line (blue curve) and Si \textsc{ii} \lm 1260 absorption line (purple curve), where \oiiiuv \lm 1666 is normalized by its peak flux, and Si \textsc{ii} \lm 1260 is normalized by the median flux in the velocity range of -250 to 250 km s$^{-1}$. No signature of outflow is present.}
    \label{fig:o3si2}
\end{figure}

\begin{figure}
\centering
\includegraphics[width=\columnwidth]{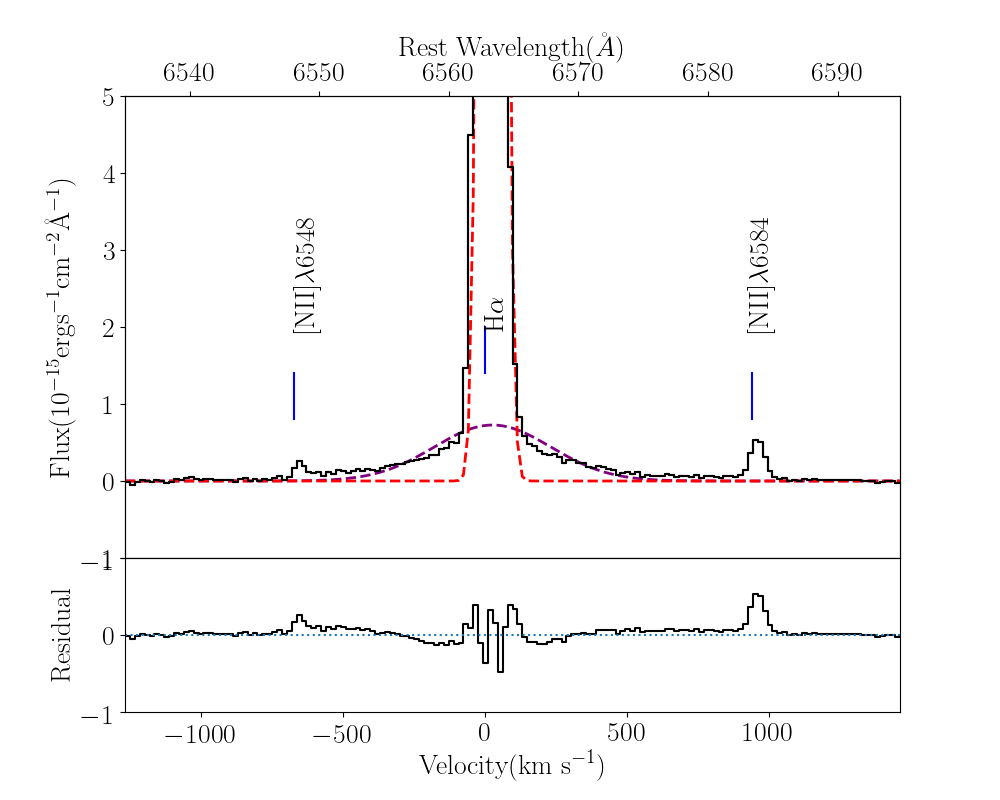}
\caption{A narrow (purple Gaussian) and a broad (red Gaussian) component is needed to reproduce the H$\alpha$ line profile, hinting towards the ISM turbulence  within Pox 186.} 
\label{fig:Ha}
\end{figure}

\begin{figure}
    \centering
    \includegraphics[width=\columnwidth]{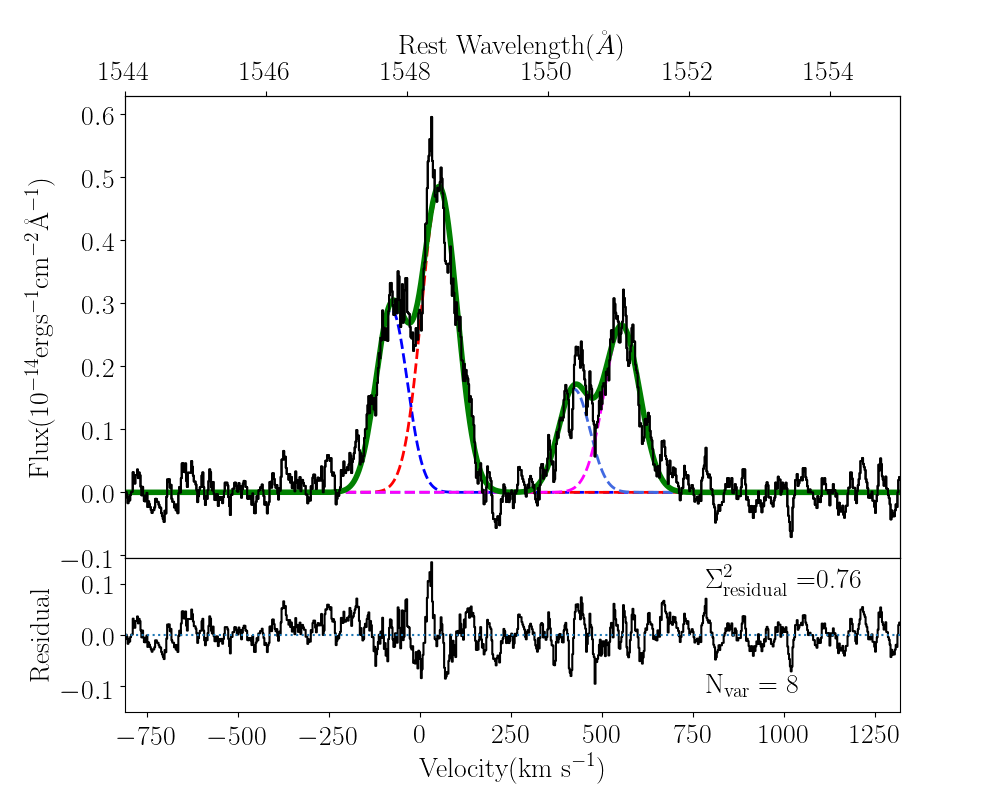}
    \includegraphics[width=\columnwidth]{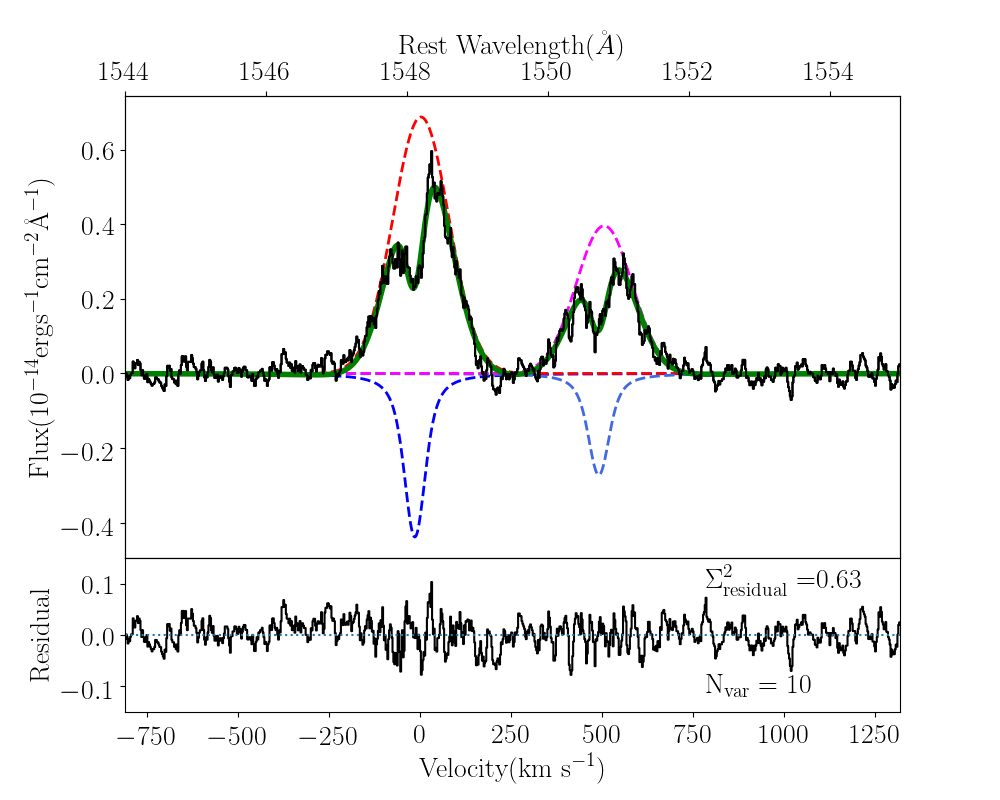}
    \caption{We model the resonantly scattered double-peaked \civ\lmlm1548,1550 doublet according to two possible scenarios: (1) Purely emission without any absorbing foreground interstellar medium in the line of sight (left-hand panel), for which we model the \civ~ doublet via multi-component Gaussian fits to identify the peaks of the blue-shifted (dashed blue curve) and the red-shifted (dashed red) components. (2) Nebular emission along with interstellar absorption (right-hand panel), where we model the \civ~ nebular emission via single Gaussian (dashed red curve) and the interstellar absorption via Voigt profile (dashed blue curve). On both panels, the overall best fit is given by the solid green line.}
    \label{fig:CIV velocity}
\end{figure}

\begin{figure*}
    \centering
    \includegraphics[width=0.8\columnwidth]{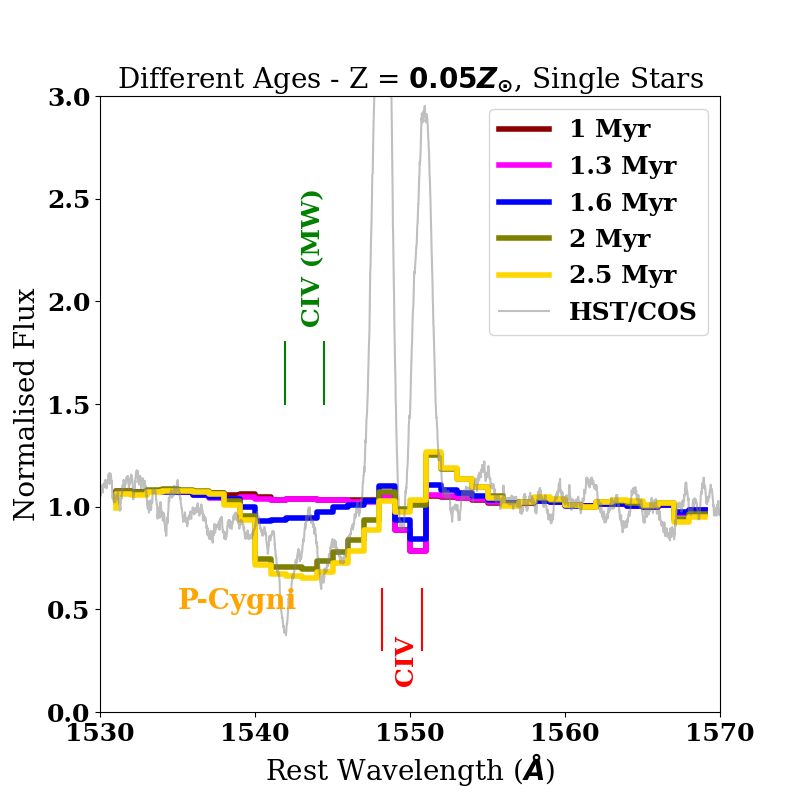}
    \includegraphics[width=0.8\columnwidth]{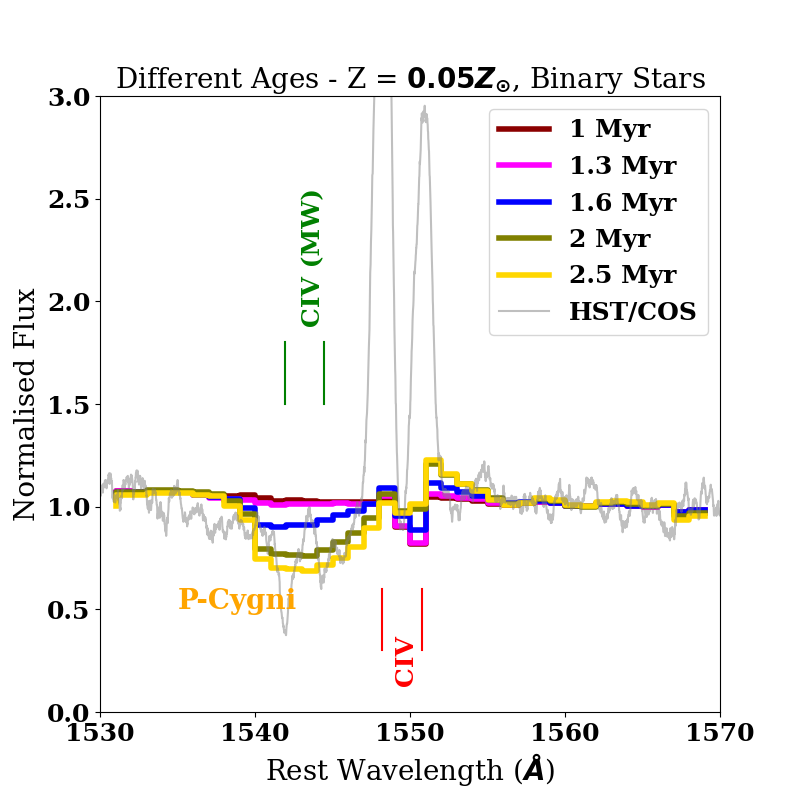}
    \caption{Comparison of observed \civ ~P-Cygni feature (grey spectrum) within Pox 186 normalized by the continuum level between 1560--1570 \AA~ with the \textsc{\textsc{bpass}} models at Z=0.05Z$_{\odot}$ comprising single (left-hand panel) and binary stars (right-hand panel) at different ages, i.e., 1 Myr (brown), 1.3 Myr (magenta), 1.6 Myr (blue), 2 Myr (olive) and 2.5 Myr (yellow). The location of \civ~ emission and absorption from Pox 186 are marked in red. The \civ~ absorption originating from MW is marked in green. }
    \label{fig:pcygni CIV}
\end{figure*}

\subsection{Line profiles of carbon lines}
\label{sec:lineprofiles}
\indent Figure \ref{fig:overlapping} (left-hand panel) shows that \civ~line profile is broadened with respect to \oiiiuv~line profile, which appears to be caused by collisional excitation. \citet{Berg2019} observed similar behaviour in a couple of local metal-poor galaxies and attributed this to the resonant scattering nature of \civ. However, for Pox 186, we find that \ciii ~line is also broadened with  respect to \oiiiuv ~(Figure \ref{fig:overlapping}, right-hand panel). Given that the \ciii ~is not reported to come from resonant scattering, we  rule out resonant scattering as the cause of broadening in carbon lines.

\indent It is unlikely that outflows could cause broadening in carbon emission lines because outflows would lead to broadening in all emission lines in the same way, which would result in similar line profiles (i.e., including \oiiiuv). Still, we explore the outflows signatures in Figure \ref{fig:o3si2}, where we overplot the \oiii ~emission line (normalized by its peak flux) along with Si \textsc{ii} \lm 1260 (normalized by the median flux within the local spectral region). Both \oiiiuv~ and Si \textsc{ii} \lm 1260 lines are centered at $\sim$ 0 km s$^{-1}$, showing no signatures of outflow. However, a hint of ionized gas suffering turbulence is present in the line profile of \ha shown in Figure \ref{fig:Ha}, which shows the presence of a broad underlying component of H$\alpha$ along with a narrow component. 

\indent Figure \ref{fig:CIV velocity} shows the velocity profile of the resonantly scattered \civ~\lmlm1548,1550 doublet. The distinct blue and red peaks exhibited by both \civ ~emission lines are quite interesting, as no previous studies have ever resolved the two peaks in both emission lines of the \civ ~doublet, though the stronger \civ~ $\lambda$1548 has been reported to have double peaks \citep{Berg2019}. The origin of double-peaked \civ~ is difficult to understand because \civ~ line profile will be impacted by the relative fraction of the gas emitting the narrow nebular emission and the foreground ISM resulting in absorption. We address two possible scenarios here: (1) a pure nebular emission with no ISM absorption  and (2) nebular emission along with ISM absorption. 

For modelling a purely nebular \civ~ (Figure \ref{fig:CIV velocity}, left panel), we fit two Gaussian components to each \civ ~emission line consisting of a blue-shifted (dashed-blue fit) and a redshifted component (dashed-red fit). The peak separation between the blue and the red peak ( V$\rm_{sep}$) is $\sim$132$\pm$2 km s$^{-1}$ for each \civ~ line, which is $\sim$25 km s$^{-1}$ (on average) higher than those found by \citet{Berg2019} for two local dwarf irregular galaxies. We note that the redshifted component is broader than the blue-shifted component.

For modelling the second scenario comprising of nebular emission and interstellar absorption (Figure \ref{fig:CIV velocity}, right panel), we model the emission in each \civ ~line with a Gaussian profile (dashed-blue fit) and interstellar absorption via the Voigt profile (dashed-red fit). 
The Voigt profile corresponding to the ISM absorption is narrower compared to the broad Gaussian profile, pointing towards a lower fraction of foreground high ionization gas along the line of sight. 

\subsection{Stellar Winds}
\label{sec:winds}
\indent Stellar winds originate from hot and massive stars, i.e., with masses $>$ $\sim$40 M$_{\odot}$ and temperatures $>$ 25,000 K, and lead to P Cygni-type UV line profiles, typically for the strongest lines, N $\textsc{v}$ \lm1240, Si \textsc{iv} \lm1400 and \civ \lm1550. The UV spectra of Pox 186 show a strong P-Cygni N \textsc{v} \lm1400 feature (Figure \ref{fig:uv_spectra1}, 1st row), no Si \textsc{iv} \lm1400 and a weak P-Cygni \civ\lm 1550 feature (Figure \ref{fig:uv_spectra1}, 3rd row). The weaker \civ\lm 1550 compared to N \textsc{v} \lm 1240 at lower metallicities, is indicative of a lower wind density and velocity \citep{Leitherer2001,Leitherer2010}. It is likely that the early O main-sequence stars are the main constituent of Pox 186, since these stars do not display wind effects in Si \textsc{iv} \lm1400, but in N \textsc{v} \lm1240 and \civ \lm 1550 as we find in Pox 186 spectra.

\indent Figure \ref{fig:pcygni CIV} shows the \textsc{bpass} models (v2.1, IMF slope = -2.35 and upper stellar mass-limit = 300 M$_{\odot}$) overlaid on \civ~ P-Cygni profile (normalized by the continuum at \lm$\sim$1560--1570\AA) both for the single stars (left-hand panel) and binary stars (right-hand panel) population for stellar ages varying from 1--2.5 Myr and metallicity of Z = 0.05Z$_{\odot}$. It appears that the stellar populations as young as 1.6 Myr are sufficient to reproduce the weak \civ~ P-Cygni profile. The inclusion of binary stars has no effect on the overall \civ~ profile, as the binary stellar population becomes more important only at later ages (3--5 Myr) \citep{Eldridge2020}.     

\indent The blue-ward absorption in the \nv P-Cygni is blended with the Ly$\alpha$ absorption from Pox 186 which is itself blended with the Ly$\alpha$ absorption from the MW. Before comparing the \nv P-Cygni with the stellar population syntheses models such as \textsc{bpass}, it is necessary to model and remove the Ly$\alpha$ absorption from Pox 186 and the MW, which requires careful modelling of the stellar continuum. We will present the detailed modelling of these components in a follow-up paper.

\subsection{Implications for JWST+ALMA studies of early galaxies}

\subsubsection{Apparent absence of \lya}
\label{sec:lya}
\indent The absence of resonantly-scattered \lya~ emission in spite of strong \ciii~ and \civ~ emissions is worth-noting (Figure \ref{fig:uv_spectra1}, upper-panel). A positive correlation is suggested/expected between EW(\ciii) and EW(\lya) by a few studies \citep[e.g.,][]{Stark2014}, though \citet{Rigby2015} suggest that a positive correlation exists for strong emitters with EW(\ciii)$>$5\AA~ and EW(\lya) $>$ 50\AA, with correlation getting weaker for weaker emitters. Given that Pox 186 shows the highest EW(\ciii) detected in the local Universe so far, we expect at least some \lya~ emission. Similarly, \citet{Berg2019} propose that the double-peaked structure of resonantly-scattered \civ ~emission could be associated with a double peak in \lya~ as well. Moreover, Pox186 shows log $\mathcal{U}$ = -2.4 $\pm$ 0.4 by  or log (q/cm s$^{-1}$) = 8.1 $\pm$ 0.4, which lies in the range of z$\sim$2-3 Ly$\alpha$ emitters \citep{Nakajima2014} further suggesting that Pox 186 could be a \lya~ emitter.
To investigate this further, we inspected the UV spectrum of Pox 186 taken with Space Telescope Imaging Spectrograph (STIS/HST) dataset (PI: Corbin, PID: 8333) which indeed shows \lya ~emission (Figure \ref{fig:STIS}). We note that the STIS and our COS pointings are offset by $\sim$ 2 arcsec ($\sim$ 168 pc), which indicates that the region emitting \lya~ is not entirely overlapping with that emitting ionized carbon, and lies at the outskirts of Pox 186 probably because of \lya~ escaping due to the concentrated feedback from the star-formation \citep{Heckman2011}. Only a deep spatial map of \lya~ can help identify any potential \lya~ emission within Pox 186. Moreover, a statistically significant sample of galaxies such as Pox 186 is required to establish any spatial offset between \lya~ emission and carbon emission. Such spatial offsets ($\sim$168 pc) between \lya~ emission and \ciii~ or \civ~ emission may not be probeable/distinguishable within the reionization era galaxies, simply because of the angular resolution of the existing instruments. So, even if originating from different regions of galaxies, UV carbon emission lines, particularly the stronger \ciii~ line, might still be a good indicator of \lya~ emission emerging from the ISM of galaxies even when \lya~ is unavailable at redshifts $>$ 6 caused by a significantly large IGM neutral fraction \citep[e.g.,][]{Fan2006}.            

\subsubsection{Lyman Continuum escape fraction}
\label{sec:escape}

\begin{figure}
    \centering
    \includegraphics[width=0.5\textwidth]{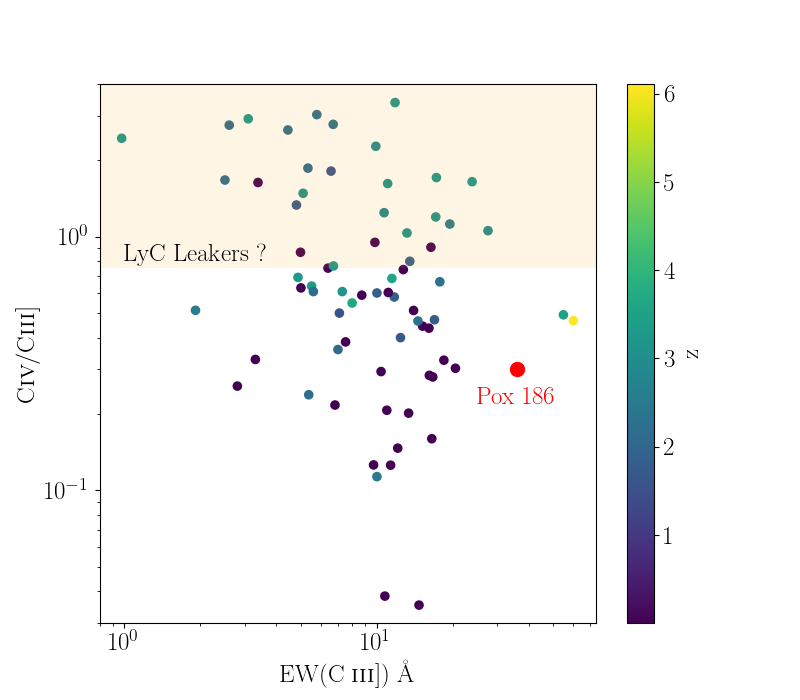}
    \caption{\civ/\ciii~ versus EW(\ciii) for Pox 186 (red point) and for published data at different redshifts presented in \citet{Schmidt2021}. The orange band represents the \civ/\ciii $>$ 0.75, which is suggested by \citet{Schaerer2022} as strong continuum leakers. Error bars are simply not shown for clarity.}
    \label{fig:Schmidt2021}
\end{figure}

\indent Figure \ref{fig:Schmidt2021} shows \civ/\ciii~ versus EW(\ciii) for Pox 186 (red point), along with measurements for galaxies at different redshifts compiled in \citet{Schmidt2021}. \citet{Schaerer2022} finds that a \civ/\ciii $>$ 0.75 (shaded orange region) is characteristic of strong Lyman Continuum (LyC) leakers, i.e. galaxies with LyC escape fraction \fesc $>$ 0.1.  Pox 186 exhibits \civ/\ciii = 0.30$\pm$0.04, which is much smaller than that exhibited by the strong LyC leakers known so far in the nearby Universe. 

\citet{Chisholm2022} suggests that strong LyC leakers have lower values of observed UV continuum slope ($\beta$). We estimate \fesc $\sim$ 10$^{-4}$ from the $\beta$ slope. 

\indent A negligible \fesc~ in Pox 186 is consistent with the absence of \lya~ for the COS pointing, however, it is inconsistent with an extreme \oiii ~\SI{88}{\micro\meter}/\cii \SI{157}{\micro\meter} shown by the Herschel/PACS maps of Pox 186. \citet{Katz2023} demonstrates that \cii~deficit (as in Pox 186) is a good indicator of LyC leakers. We note that the resolution of the PACS data is much worse ($\gtrsim$ 9\arcsec \footnote{For reference, the beam size of PACS spectrometer is $\sim$9\arcsec and 12\arcsec at ~\SI{60}{\micro\meter} and \SI{150}{\micro\meter}, respectively.}) than that of our UV and optical data. It is possible that there might be a spatial offset between the region resulting in high  \oiii ~\SI{88}{\micro\meter}/\cii \SI{157}{\micro\meter} and region emitting bright UV lines. 

\indent \citet[][]{Izotov2018} reports an increase in \fesc~ with an increase in \oiii/\oii though with a large scatter.  We could not estimate \oiii/\oii of Pox 186 either at the spatially-resolved data or using the integrated spectrum coinciding with the COS aperture as \oii\lm\lm3727,3729 doublet is undetected. \citet{Guseva2004} present the emission line fluxes of Pox 186 within slits of 1\arcsec$\times$3.2\arcsec and 2\arcsec$\times$6\arcsec. Using those line fluxes, we estimate \oiii/\oii = 18-22 for Pox 186 indicating a high \fesc. Moreover, \citet{Ramambason2022} reports a 40\% escape fraction of ionizing photons for Pox 186 by using a suite of IR lines.

\indent Figure \ref{fig:Schmidt2021} also shows a couple of reionization era galaxies which have \civ/\ciii ~ below the threshold proposed by \citet{Schaerer2022}, and close to Pox 186. This raises two questions: (1) Is \civ/\ciii~ or $\beta$ slope a good enough predictor for LyC leakers? (2) Were there EoR galaxies which did not contribute to the reionization of the Universe? Such questions can be addressed by making the simultaneous use of UV and FIR observations of large samples of EoR galaxies as well as their local analogues. It would be crucial to do follow-up UV+optical observations via JWST of EoR  galaxies, which show extreme IR properties via ALMA, and vice versa.  


\section{Summary}
\label{sec:summary}

\indent We investigate the ionized carbon within a local dwarf galaxy Pox 186 using the HST/COS UV data complemented by the GMOS-IFU optical data and Herschel FIR data. Our main results are summarized as follows:
\begin{enumerate}
    \item Using the HST/COS UV emission lines, we measure a redshift z = 0.0040705 $\pm$ 0.000013 for Pox 186. This corresponds to a luminosity distance of 17.5 Mpc assuming a flat $\Lambda$CDM cosmology or 12.6 Mpc assuming Cosmicflows-3.

    \item The COS/UV data reveals very high EW of carbon emission lines, i.e. EW(\ciii) = 35.85$\pm$0.73\AA~ and EW(\civ) = 7.75$\pm$0.28\AA.

    \item We explore several scenarios to explore the high EW of carbon lines, including a high effective temperature, higher than average carbon-to-oxygen ratio for a given gas-phase metallicity, photoionization \textsc{cloudy} models including binary stars, top-heavy IMF and nebular $\upalpha/Fe$ enhancement and in-homogeneous dust-distribution. The photoionization models could not simultaneously reproduce all the observables irrespective of dust-reddening, which could be due to the simplistic assumptions of the model parameters.

    \item The \ciii~ and \civ ~ lines also show broadening with respect to the \oiiiuv~ emission lines though the cause of this broadening remains unknown. We rule out outflows causing broad carbon emission lines as no outflow signatures are found in the velocity profiles of \oiiiuv~ and  Si \textsc{ii} lines. 

    \item Optical integrated spectrum coinciding with COS aperture, shows a broad and faint underlying component in \ha ~along with a narrow component. Ruling out outflows on the basis of UV data, the \ha~ velocity profile indicates a turbulent ISM.

    \item The \civ~ doublet shows clearly distinct double peaks for each of the two emission lines, which can be explained via two scenarios, such as pure emission with no absorbing foreground ISM or nebular emission along with a little absorbing ISM in the foreground.

    \item The high EW(\ciii) and log $\mathcal{U}$ = -2.4 $\pm$ 0.4 suggests a high EW(\lya) for Pox 186; however, COS spectra do not show any signature of \lya~ though a spatially-offset STIS spectrum does show \lya~ emission.

    \item We report an observed UV continuum slope $\beta$ = --0.36$\pm$0.04 which corresponds to \fesc$\sim$10$^{-4}$,  indicating that Pox 186 is not a LyC leaker. \civ/\ciii~ is also below the threshold for LyC leakers suggested by \citet{Schaerer2022}. This is in contrast with the extreme \oiii/\cii FIR line ratio, 40\% escape fraction \citep{Ramambason2022} and the high \oiii/\oii values from the literature. This raises questions on the potential use of $\beta$ or \civ/\ciii~ as tracers of LyC leakers.
    
\end{enumerate}

This work shows that the extreme IR \oiii/\cii emission line ratios could correspond to extreme UV properties such as high EW of carbon lines (\ciii~ and \civ), high carbon-to-oxygen ratio, broadened emission carbon line profiles and double-peak within the resonant carbon line doublet, \civ. However, the apparent absence of \lya~ emission and negligible LyC escape fraction (as estimated from UV slope and \civ/\ciii ratio) within a dwarf galaxy with such extreme UV and IR properties are puzzling. This requires a similar investigation on a larger sample of similar galaxies with UV+FIR data. The combination of HST and Herschel data for the local Universe, and JWST and ALMA for the reionization era Universe are crucial in carrying out such studies and understanding the similarities and differences between the EoR galaxies and their local analogues.    

\section*{Acknowledgements}

NK thanks Joe Hunkeler for their help in setting up the new Gemini \textsc{iraf} reduction pipeline and debugging the issues with the pipeline. NK thanks Elizabeth Stanway for clarifying information on \textsc{bpass} models. RS acknowledges financial support from the UK Science and Technology Facilities Council (STFC). JW acknowledges support from the ERC Advanced Grant 695671, ``QUENCH'', and the Fondation MERAC.  
\indent This research is based on observations made with the NASA/ESA Hubble Space Telescope obtained from the Space Telescope Science Institute, which is operated by the Association of Universities for Research in Astronomy, Inc., under NASA contract NAS 5–26555. These observations are associated with programs GO 16071 and 16445. 

\indent This work is further partially based on observations obtained at the Gemini Observatory, which is operated by the Association of Universities for Research in Astronomy, Inc., under a cooperative agreement with the NSF on behalf of the Gemini partnership: the National Science Foundation (United States), the Science and Technology Facilities Council (United Kingdom), the National Research Council (Canada), CONICYT (Chile), the Australian Research Council
(Australia), Ministério da Ciência e Tecnologia (Brazil) and SECYT (Argentina)

\indent This research has made use of NASA's Astrophysics Data System Bibliographic Services'; SAOImage DS9, developed by Smithsonian Astrophysical Observatory’; Astropy, a community-developed core PYTHON package for Astronomy (Astropy Collaboration et al. 2013); matplotlib \citep{Hunter2007} and numpy \citep{Harris2020,Van2011}. 

\indent This research has made use of the Spanish Virtual Observatory (https://svo.cab.inta-csic.es) project funded by MCIN/AEI/10.13039/501100011033/ through grant PID2020-112949GB-I00.
\section*{Data Availability}

The data presented in this paper are available in the Multimission Archive at the Space Telescope Science Institute (MAST) and Gemini Observatory Archive.



\bibliographystyle{mnras}
\bibliography{biblio.bib} 




\appendix

\section{Target acqusition image}
\label{app:targacq}
\begin{figure}
    \centering
    \includegraphics[width=0.45\textwidth, trim={0.5cm 0cm 0.5cm 0cm}, clip]{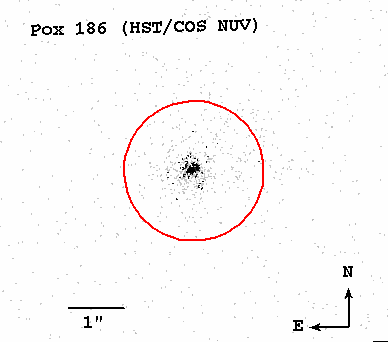}
    \caption{HST/COS NUV target acquisition image of Pox 186. The red circle denotes the 2.5 arcsec COS spectroscopic aperture and is centered at (RA, Dec): (13 25 48.641, -11 36 37.94). The compass on the bottom-right of the figure shows North and East on the HST image. At the luminosity distance of this galaxy (i.e. 17.5Mpc), 1 arcsec corresponds to 84 pc.}
    \label{fig:NUV/Acq}
\end{figure}

Figure \ref{fig:NUV/Acq} shows one of the two NUV target acquisition images taken with taking the spectroscopic COS observations.

\section{Optical flux scaling and systematic uncertainties}
\label{app:fluxscale}

Note that the flux calibration obtained from the spectrophotometric standard star is relative and not absolute. To determine the optical emission line fluxes accurately, we follow the procedure described below. We first compare the ground-based Gemini/optical spectroscopic fluxes with the space-based archival HST imaging taken with the F658N filter. To do so, we extract the flux within an aperture of 1.25\arcsec radius (matching the COS aperture) on the WFC3/F658N image, subtract a background and apply the aperture correction. This value is then multiplied with the inverse sensitivity and the FWHM of the filter (estimated from synthetic photometry package \textsc{synphot}) to get the total flux within the aperture in the cgs units. We note that the width of an HST/WFC3 filter can be defined in several ways; however, we chose to use FWHM following the methodology of \citet{Laigle2016} for generating large photometric catalogues. To get the WFC3/F658N equivalent from Gemini optical data, we multiply the COS-matched Gemini spectrum with the transmission curve of the WFC3/F658N filter available from the SVO Filter Profile Service \citep{Rodrigo2012, Rodrigo2020}. We integrate the resultant spectrum to get the HST equivalent of Gemini flux. This gives us a scaling factor of $\sim$2.7 for the Gemini spectra taken with R831 grating at $\sim$ 6900\AA.  

Note that we have three sets of Gemini spectra taken B600 grating at $\sim$4650\AA~ and R831 grating at 6900\AA~ and 8900\AA. The Gemini spectra taken with B600/4650\AA and R831/8900\AA~ could in principle be scaled with respect to the scaled Gemini spectra taken with R831/6900\AA, however, the continuum is too faint to do such an exercise. So, we estimate the E(B-V) from the unscaled H$\gamma$/H$\beta$ ratio from the B600/4650\AA, and then use it with the scaled Ha line flux to estimate the scaling factor for H$\beta$ and hence the corresponding spectra. A similar procedure based on E(B-V) is followed where we predict the H I P10 line flux obtained from the R831/8900\AA~ spectra, thus determining the scaling factor of these spectra. 

These scaling factors give us H$\beta$ flux value in reasonable agreement with that derived by \citet{Guseva2004} for different-sized slits.

\indent To determine the systematic uncertainties on flux calibration, we compare the integrated Gemini/IFU spectra of the standard photometric star (HZ44) with that available from CALSPEC, which gives us a systematic uncertainty of ~50\%.

\section{Optical Emission line flux maps}
\label{app:maps}

\begin{figure*}
\centering
\includegraphics[width=0.33\textwidth, trim={1.8cm 0.5cm 0.2cm 0.5cm}, clip]{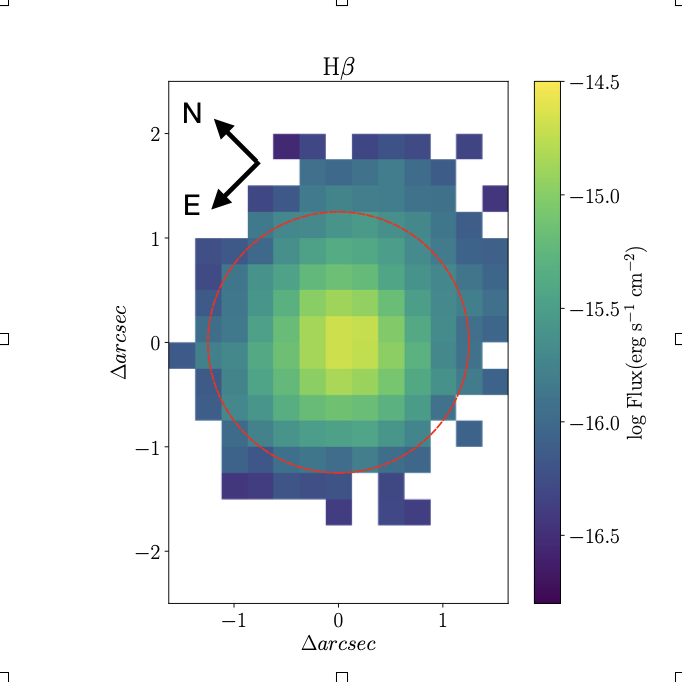}
\includegraphics[width=0.33\textwidth, trim={3.0cm 0.5cm 0cm 0cm}, clip]{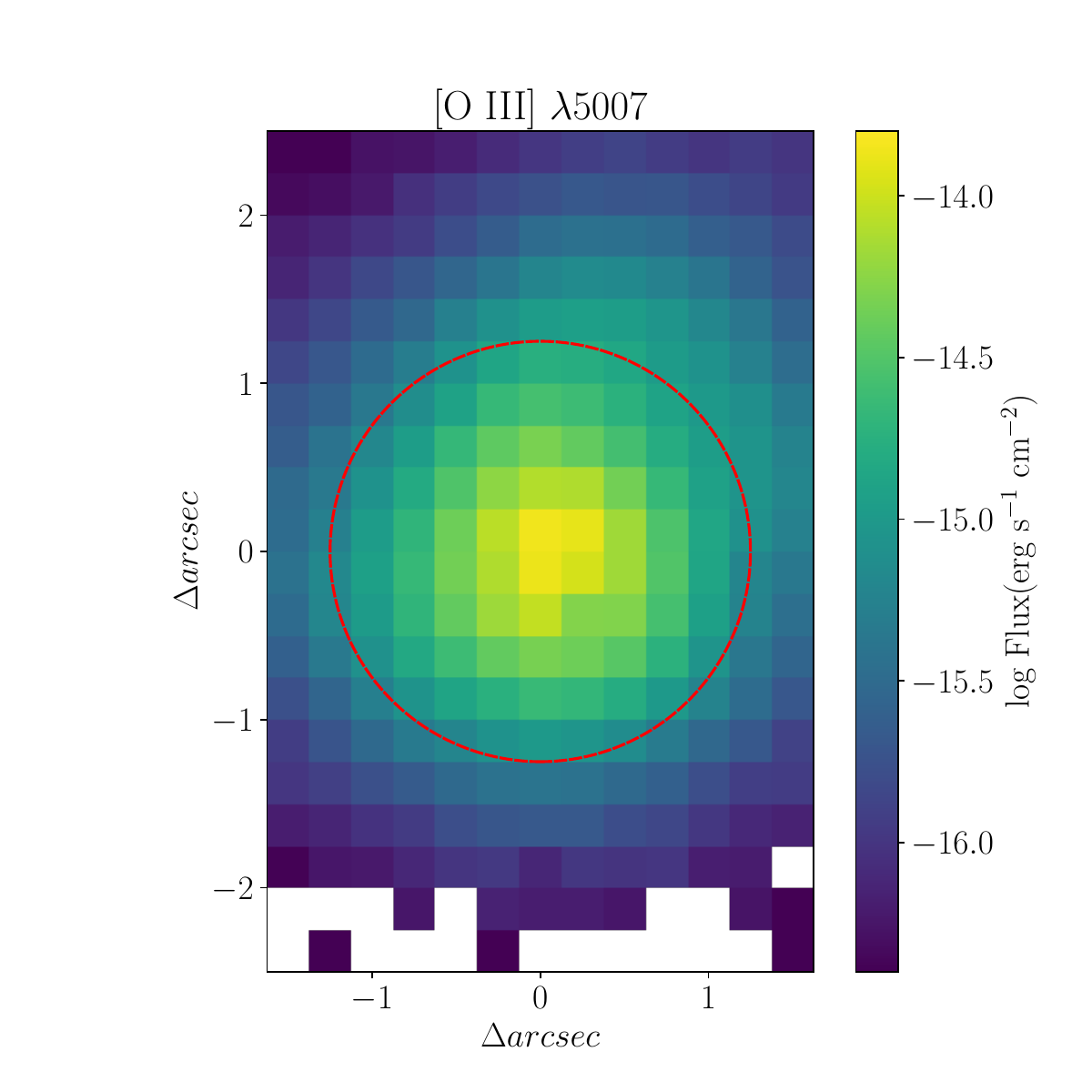}
\includegraphics[width=0.33\textwidth, trim={3.0cm 0.5cm 0cm 0cm}, clip]{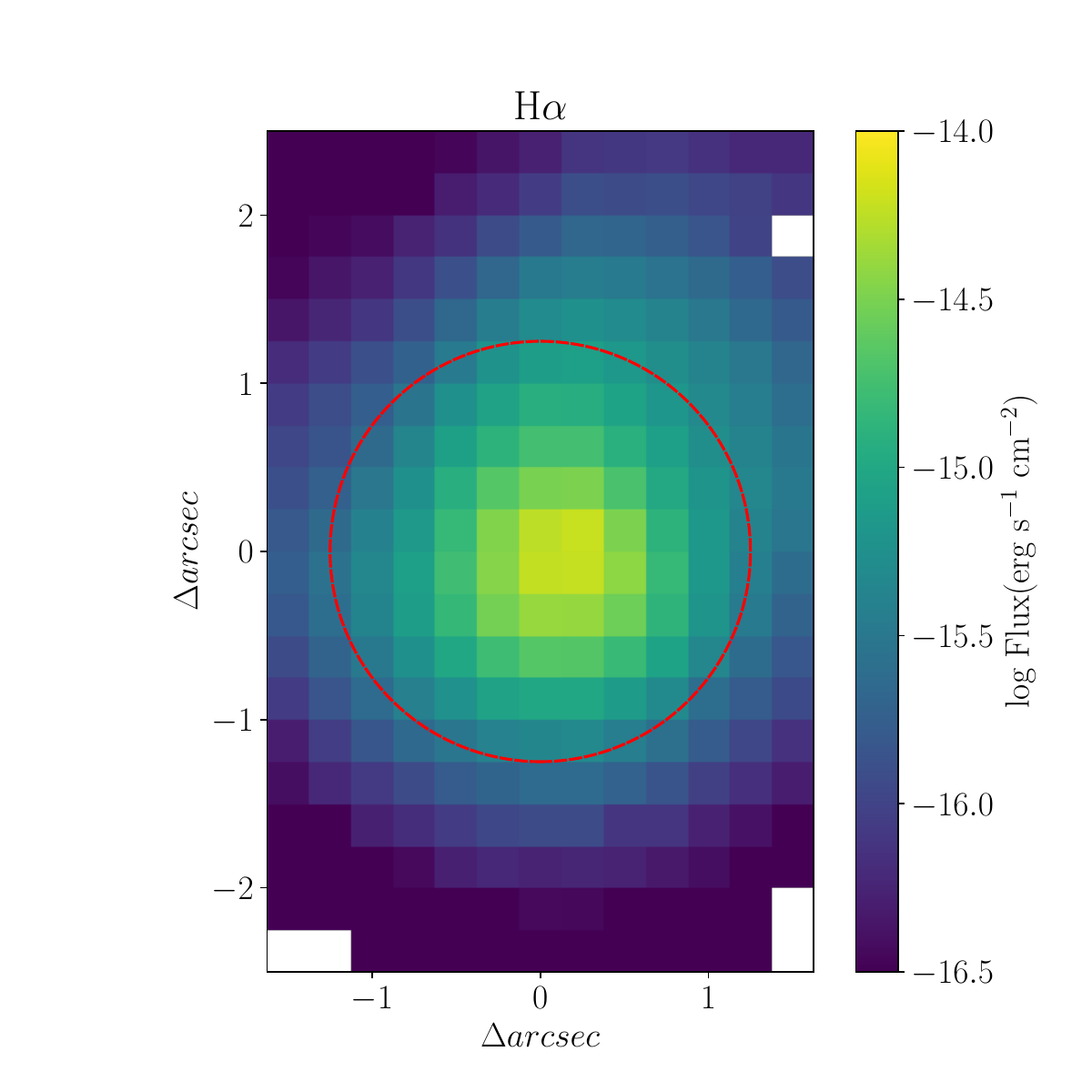}
\includegraphics[width=0.33\textwidth, trim={3.0cm 0.5cm 0cm 0cm}, clip]{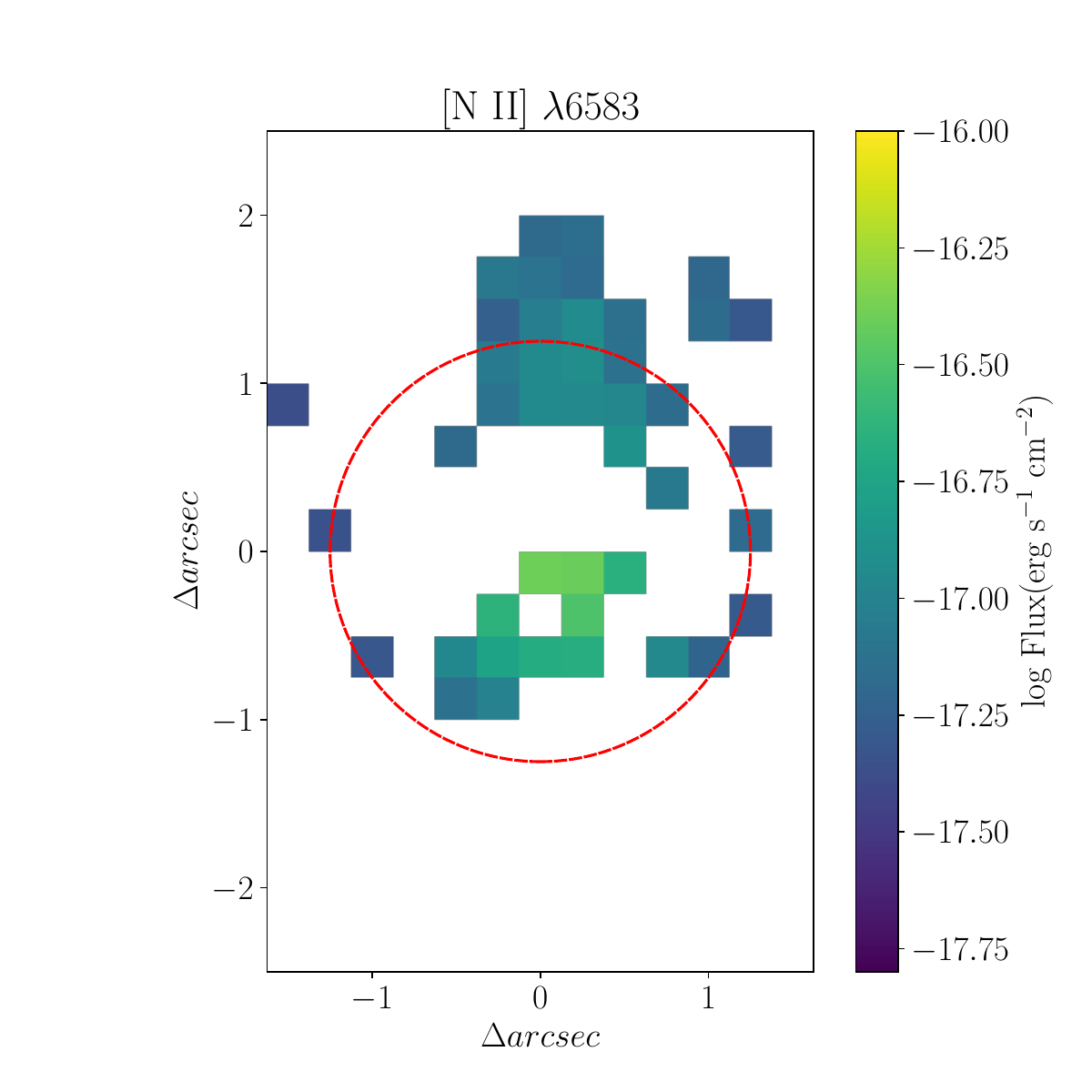}
\includegraphics[width=0.33\textwidth, trim={3.0cm 0.5cm 0cm 0cm}, clip]{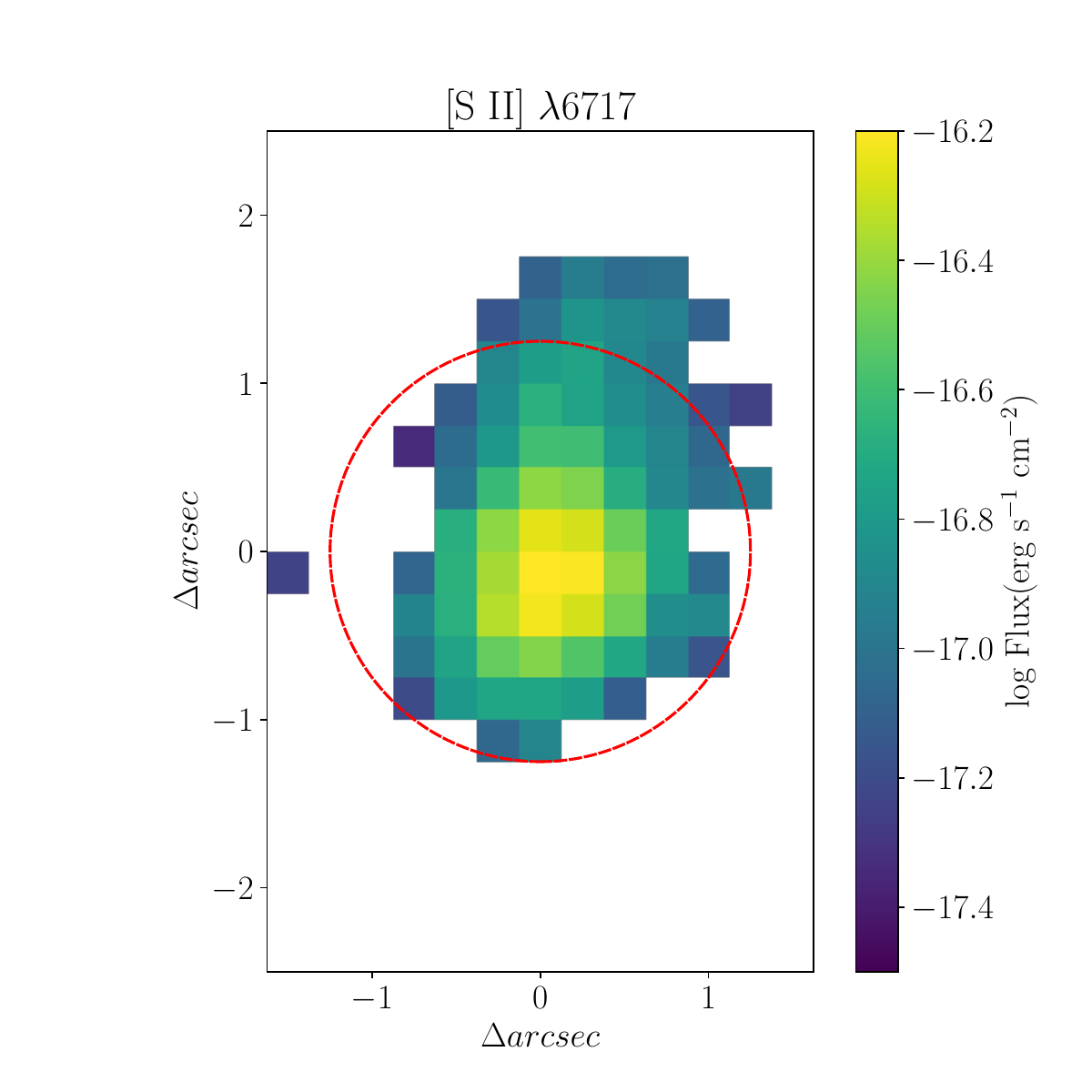}
\includegraphics[width=0.33\textwidth,trim={3.0cm 0.5cm 0cm 0cm}, clip]{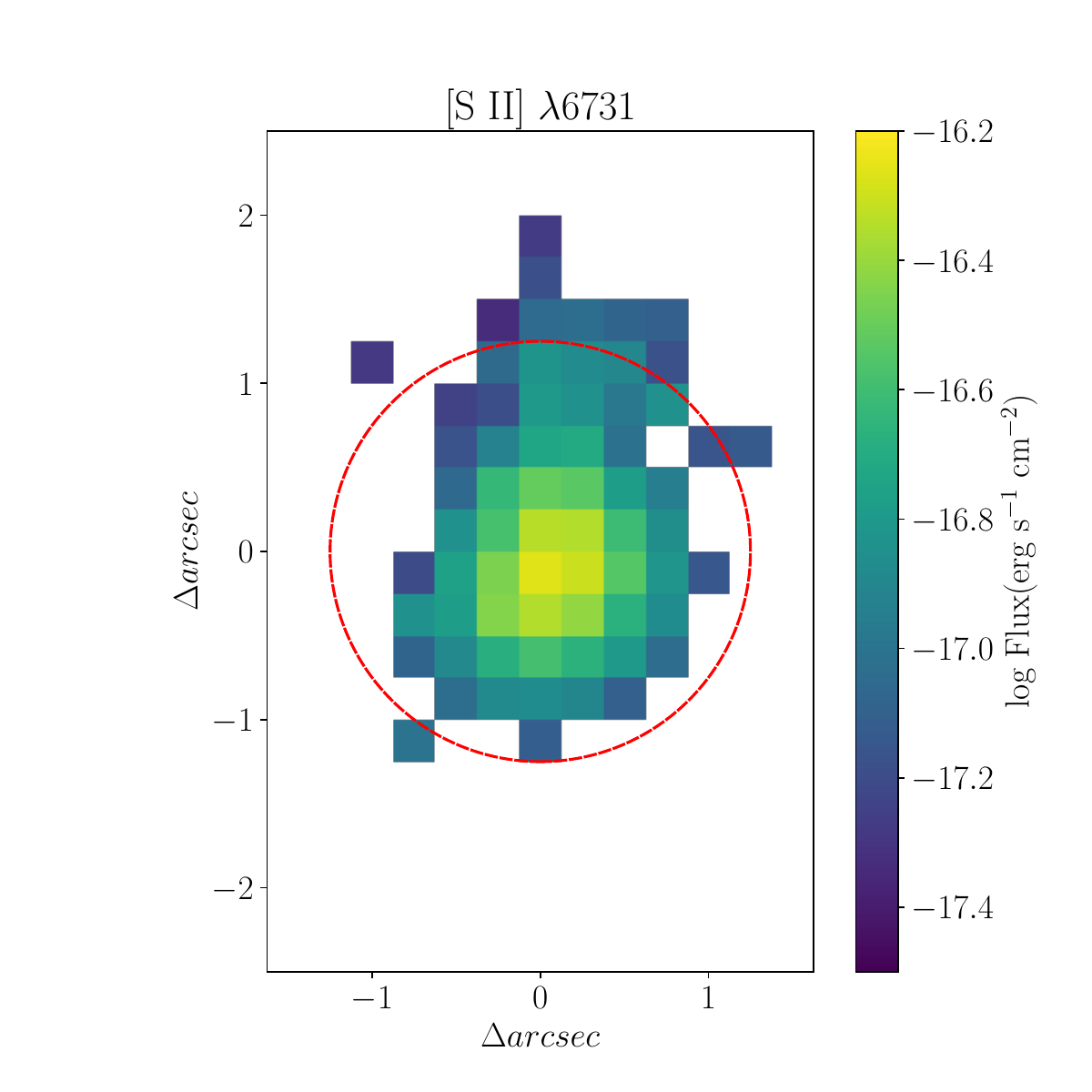}
\caption{Optical emission line flux maps of Pox186 obtained from GMOS-N IFU taken as part of program GN-2020A-FT-105. The orientation of the maps is indicated by the compass on the \hb flux map. The white pixels indicate the regions where S/N < 3. The red circle indicates the location of COS aperture.}
\label{fig:a_flux}
\end{figure*}

\begin{figure*}
\centering
\includegraphics[width=0.33\textwidth, trim={3.0cm 0.5cm 0cm 0cm}, clip]{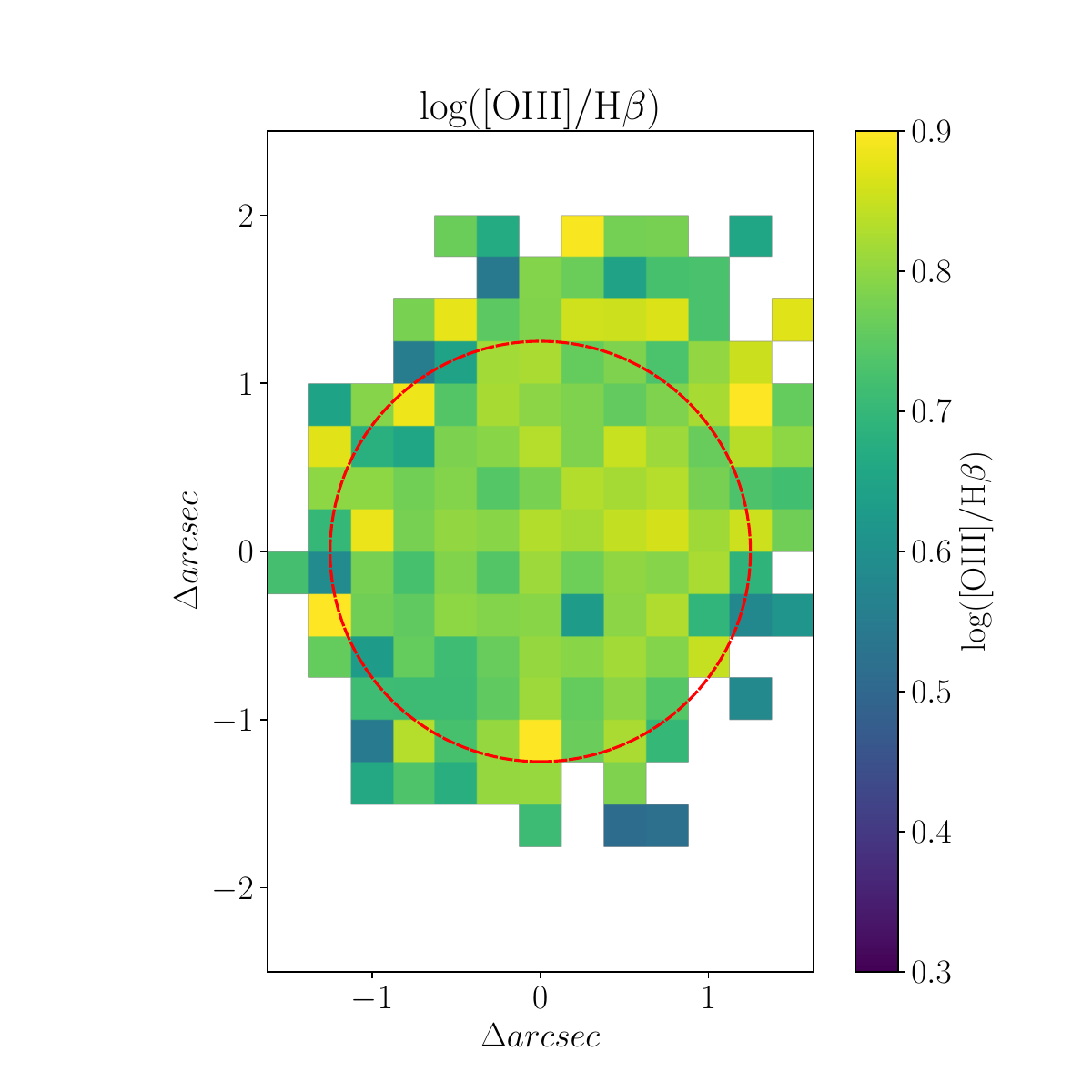}
\includegraphics[width=0.33\textwidth, trim={3.0cm 0.5cm 0cm 0cm}, clip]{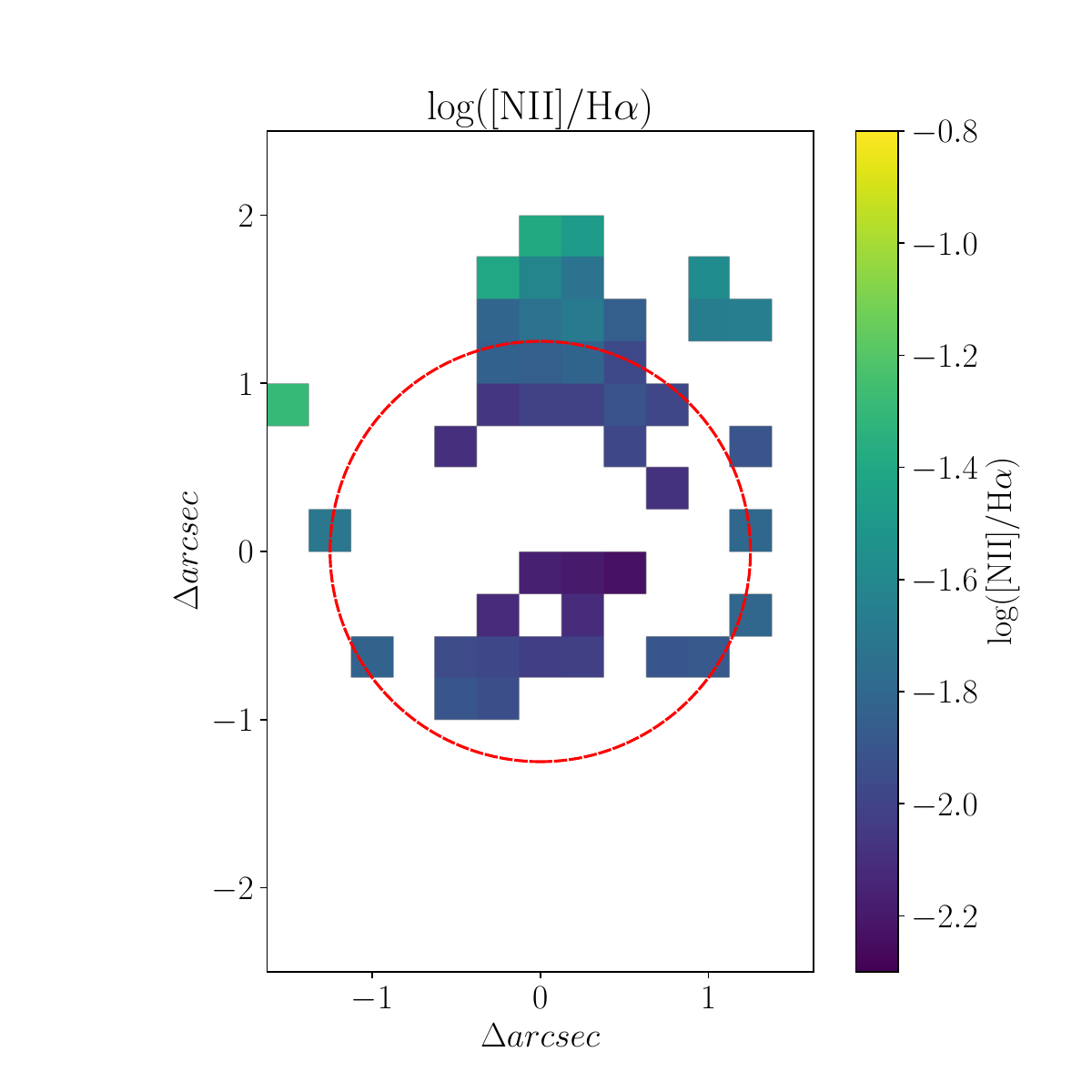}
\includegraphics[width=0.33\textwidth, trim={3.0cm 0.5cm 0cm 0cm}, clip]{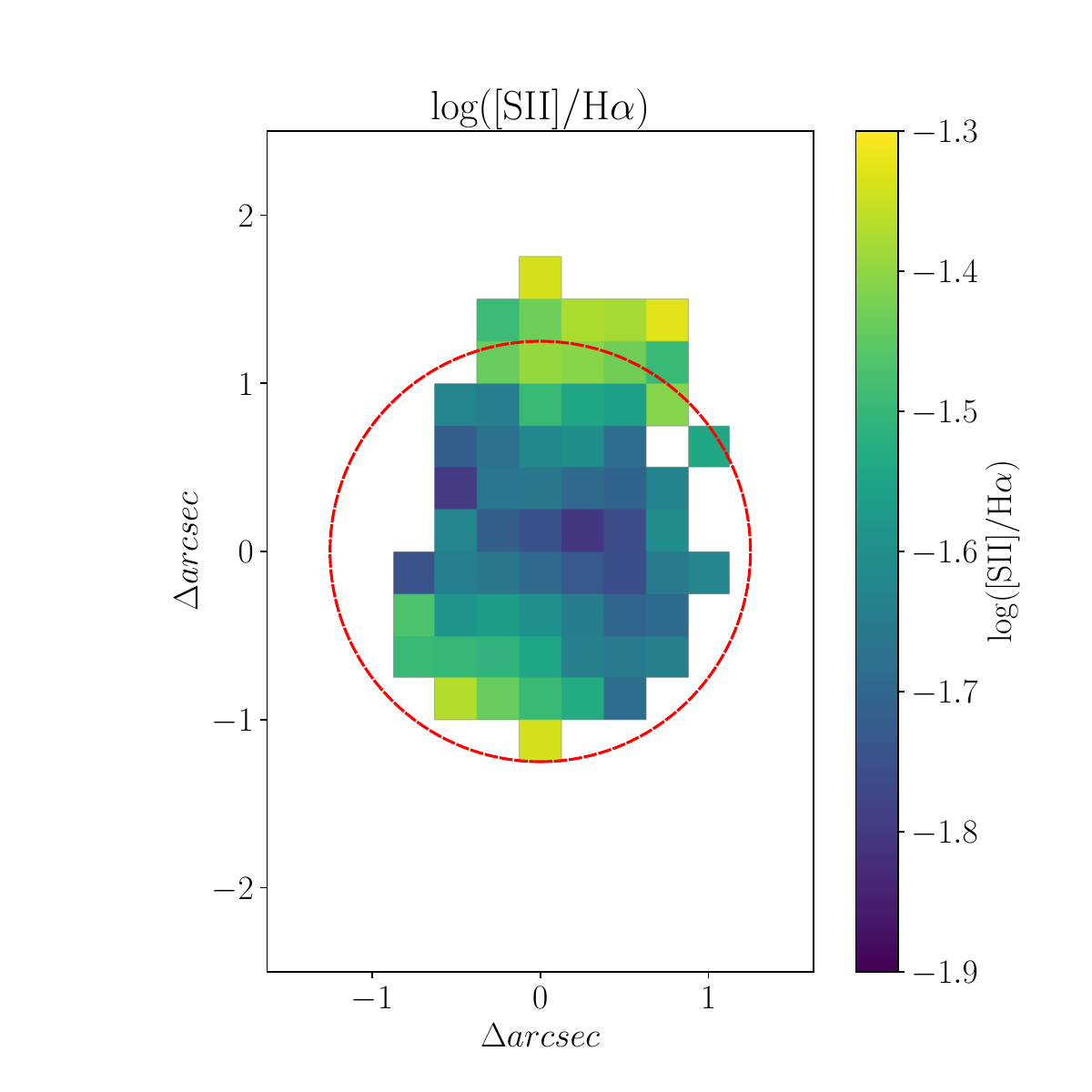}
\caption{Maps of the optical emission line ratios created from using the flux maps shown in Figure \ref{fig:a_flux}: \oiii/\hb (left-hand panel), \nii/\ha (middle panel) and \sii/\ha (right-hand panel). The white pixels indicate the regions where S/N < 3. The red circle indicates the location of COS aperture.}
\label{fig:bpt_maps}
\end{figure*}

\begin{figure}
\centering
\includegraphics[width=0.33\textwidth, trim={1.5cm 0.5cm 0.2cm 0.5cm}, clip]{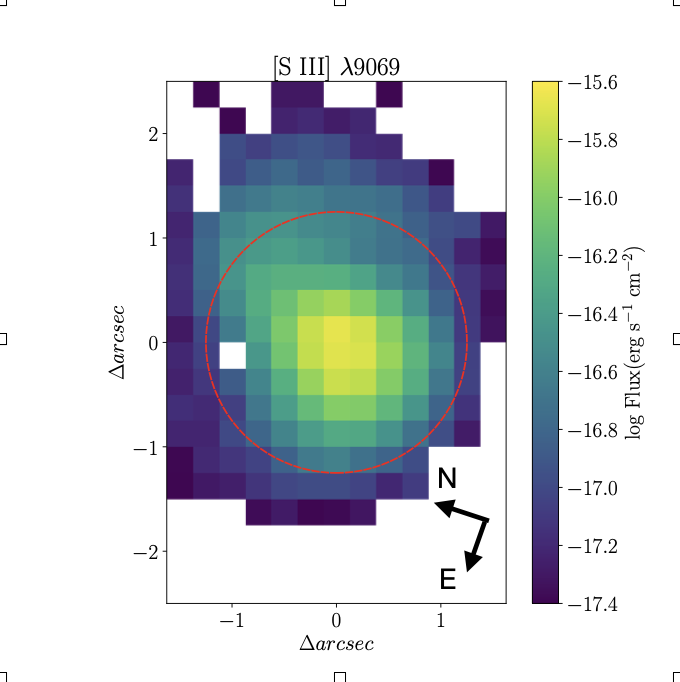}
\caption{\siii \lm 9069 emission line flux map obtained from GMOS-N IFU data taken as part of the program GN-2021-FT-111. The orientation of the map is indicated by the compass on the lower-left corner. The white pixels indicate the regions where S/N < 3. The red circle indicates the location of COS aperture.}
\label{fig:a_flux_nir}
\end{figure}

Figures \ref{fig:a_flux} and \ref{fig:a_flux_nir} show the emission line flux maps obtained by fitting a single Gaussian to each optical line. The orientation of the FOV is different in the two figures because each set of Figures here is obtained from different observing programs with different position angle (see Figure \ref{fig:fov}). A compass is shown in each figure to show the orientation. Figures \ref{fig:bpt_maps} show the maps of optical emission line ratios used in the classical emission line ratio diagrams (or the so-called BPT diagrams). 

\section{BPASS models results with IMF slope -2.35 and without \texorpdfstring{$\upalpha$/Fe}{} enhancement}
\label{app:bpass}

\begin{figure}
    \centering
    \includegraphics[width=0.45\textwidth]{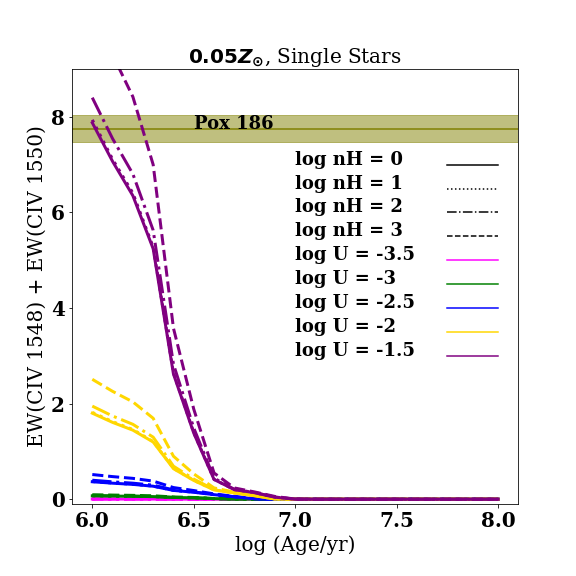}
    \includegraphics[width=0.45\textwidth]{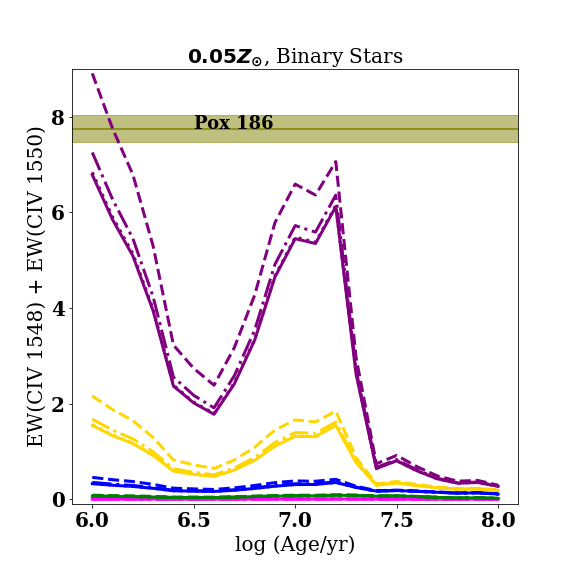}
    \caption{\textsc{bpass} models showing the variation of the equivalent width of \civ emission lines for single stars (upper panel) and binary stars (lower panel), and for a metallicity of 0.05Z$_{\odot}$. Note that we also produced similar plots for other metallicities but a metallicity of 0.05Z$_{\odot}$ could only reproduce the EW(\civ) as demonstrated in the lower panel. Five values of log $\mathcal{U}$ are considered i.e., -1.5 (purple), -2.0 (yellow), -2.5 (blue), -3.0 (green), -3.5 (magenta). Similarly, we consider four values of hydrogen densities, log n$_H$, i.e., 0 (solid curve), 1 (dotted curve), 2 (dash-dot curve), 3 (dashed curve). The observed EW of \ciii and (upper panel) \civ~ (lower) and the associated uncertainties are represented by the solid horizontal black line and the shaded olive green region. Single star models with any n$_H$, but high log $\mathcal{U}$ (=--1.5) an low metallicity (0.05 Z$_{\odot}$) can reproduce the observed EW(\civ) at ages of log(Age/yr) = 6-6.5. However, when binary stars are considered, very high hydrogen  density (log n$_{H}$ = 3), high ionization parameter (log $\mathcal{U}$ = -1.5) and low metallicity (0.05 Z$\odot$) are required to reproduce the observed EW(\civ) at ages of log(Age/yr)$\sim$6.25.}
    \label{fig:bpass_binsin}
\end{figure}

\indent We explore whether the extreme radiation field from the stellar population could cause the high EW of carbon lines by using the BPASS models without $\upalpha$/Fe enhancement. Figure \ref{fig:bpass_binsin} shows \textsc{bpass} models (version 2.2.1, upper IMF slope of -2.35 and upper mass limit of 300 M$\rm_{\odot}$) consisting of single stars at stellar metallicity of 0.05 Z$\rm_\odot$ and age of $\sim$1 Myr are able to reproduce observed EW(\civ) for log U = -1.5 and any value of hydrogen density (n$\rm_H$). However, the same \textsc{bpass} model consisting of binary stars is able to produce the observed EW(\civ) only for hydrogen density log n$\rm_H/cm^{-3}$ = 3.  

\begin{figure}
    \centering
    \includegraphics[width=0.45\textwidth]{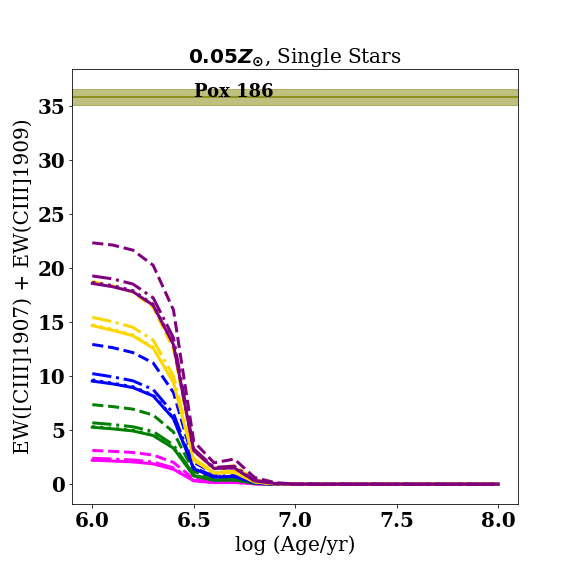}
    \includegraphics[width=0.45\textwidth]{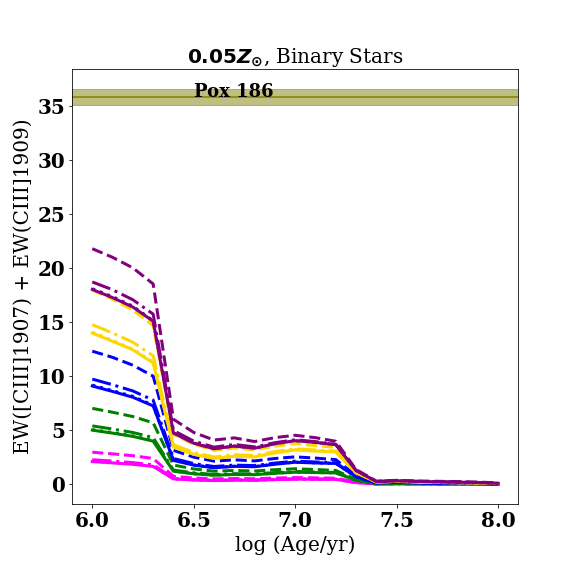}
    \caption{\textsc{bpass} models showing the variation of the equivalent width of the \ciii  emission lines for single stars (upper panel) and binary stars (lower panel), and for a metallicity of 0.05Z$_{\odot}$. Five values of log $\mathcal{U}$ are considered i.e., -1.5 (purple), -2.0 (yellow), -2.5 (blue), -3.0 (green), -3.5 (magenta). Similarly, we consider four values of hydrogen densities, log n$_H$, i.e., 0 (solid curve), 1 (dotted curve), 2 (dashdot curve), 3 (dashed curve). The observed EW of \ciii~and the associated uncertainties are represented by the solid horizontal black line and the shaded olive green region. The models sufficient to produce high \civ~ are insufficient to reproduce the high \ciii.}
    \label{fig:app_bpass_binsin}
\end{figure}

\indent Since EW(\civ\lmlm1548,1550) for Pox 186 is reproduced by the \textsc{bpass} models with a particular set of physical parameters, we use the same \textsc{bpass} models to explore whether stellar populations radiation field from the stellar population could produce the high EW(\ciii). However, neither single nor binary stars are sufficient to achieve such a high value of EW(\ciii) exhibited by Pox 186 (Figure \ref{fig:app_bpass_binsin}). Hence we discard the possibility that the extreme radiation field from the stellar population is responsible for the high EW of carbon lines.

\section{\texorpdfstring{\lya}{}~ emission within HST/STIS spectrum}
\label{app:stis}

\begin{figure*}
    \centering
    \includegraphics[width=0.9\textwidth]{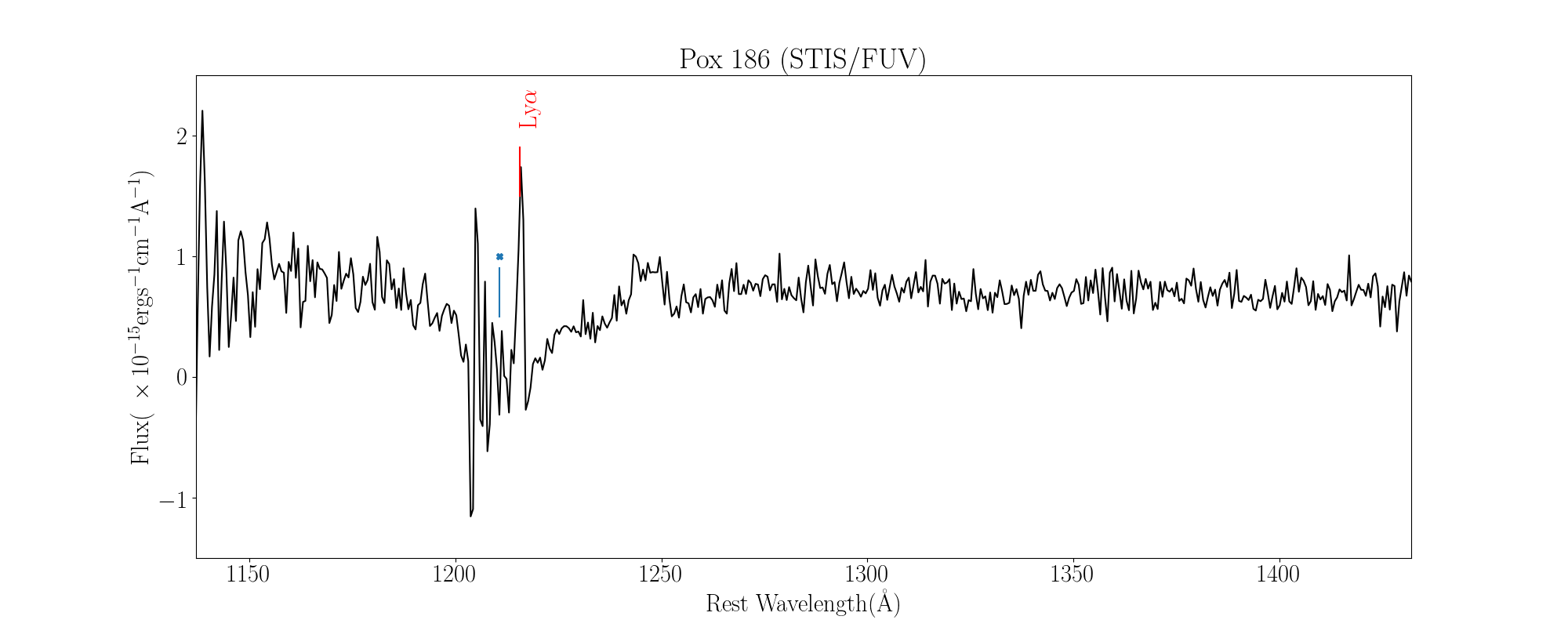}
    \caption{A noisy yet prominent \lya~ emission amidst the damped \lya absorption is detected within the STIS/FUV spectrum of Pox 186. The blue cross shows the location of geocoronal emission. }
    \label{fig:STIS}
\end{figure*}

Figure \ref{fig:STIS} shows a noisy yet prominent \lya~ emission in the UV spectrum of Pox 186 taken with Space Telescope Imaging Spectrograph (onboard HST) using the G140L grating. The aperture used for taking observation was 52$\times$0.5, which corresponds to a slit of 0.5 arcsecs, and was centred at RA and Dec of  13 25 48.50 and -11 36 37.70, respectively. This translates into a spatial offset of 2 arcsecs between the STIS and our COS pointings. 

\bsp	
\label{lastpage}
\end{document}